\documentclass[a4paper,clock]{article}
\usepackage{amssymb}
\usepackage{bm}
\usepackage{graphicx}
\usepackage{dcolumn}
\usepackage{epstopdf}

\catcode`\@=11
\def\marginnote#1{}
\newcount\hour
\newcount\minute
\newtoks\amorpm
\hour=\time\divide\hour by60
\minute=\time{\multiply\hour by60 \global\advance\minute
by-\hour}\edef\standardtime{{\ifnum\hour<12
\global\amorpm={am}%
        \else\global\amorpm={pm}\advance\hour by-12 \fi
        \ifnum\hour=0 \hour=12 \fi
        \number\hour:\ifnum\minute<10
0\fi\number\minute\the\amorpm}}
\edef\militarytime{\number\hour:\ifnum\minute<10
0\fi\number\minute}

\def\draftlabel#1{{\@bsphack\if@filesw {\let\thepage\relax
   \xdef\@gtempa{\write\@auxout{\string
      \newlabel{#1}{{\@currentlabel}{\thepage}}}}}\@gtempa
   \if@nobreak \ifvmode\nobreak\fi\fi\fi\@esphack}
        \gdef\@eqnlabel{#1}}
\def\@eqnlabel{}
\def\@vacuum{}
\def\draftmarginnote#1{\marginpar{\raggedright\scriptsize\tt#1}}
\def\draft{\oddsidemargin -.5truein
        \def\@oddfoot{\sl preliminary draft \hfil
        \rm\thepage\hfil\sl\today\quad\militarytime}
        \let\@evenfoot\@oddfoot \overfullrule 3pt
        \let\label=\draftlabel
        \let\marginnote=\draftmarginnote

\def\@eqnnum{(\theequation)\rlap{\kern\marginparsep\tt\@eqnlabel}%
\global\let\@eqnlabel\@vacuum}  }

\def\numberbysection{\@addtoreset{equation}{section}
        \def\theequation{\thesection.\arabic{equation}}}

\def\underline#1{\relax\ifmmode\@@underline#1\else
 $\@@underline{\hbox{#1}}$\relax\fi}

\catcode`@=12
\relax

\numberbysection

\advance \topmargin by -\headheight
\advance \topmargin by -\headsep

\evensidemargin \oddsidemargin
\def\br{\begin{eqnarray}}
\def\er{\end{eqnarray}}
\def\be{\begin{equation}}
\def\ee{\end{equation}}

\def\({\left(}
\def\){\right)}

\relax

\def\a{\alpha}

\def\b{\beta}

\def\d{\delta}
\def\D{\Delta}

\def\g{\gamma}
\def\G{\Gamma}

\def\l{\lambda}

\def\pa{\partial}

\def\tp0{\Theta_{+}^{(0)}}
\def\tm0{\Theta_{-}^{(0)}}

\def\f#1#2#3 {f^{#1#2}_{#3}}

\def\win1{{\sf w_{1+\infty}}}

\def\Win1{{\sf W_{1+\infty}}}

\def\rlx{\relax\leavevmode}
\def\inbar{\vrule height1.5ex width.4pt depth0pt}
\def\IZ{\rlx\hbox{\sf Z\kern-.4em Z}}
\def\IR{\rlx\hbox{\rm I\kern-.18em R}}
\def\IC{\rlx\hbox{\,$\inbar\kern-.3em{\rm C}$}}
\def\IN{\rlx\hbox{\rm I\kern-.18em N}}
\def\IO{\rlx\hbox{\,$\inbar\kern-.3em{\rm O}$}}
\def\IP{\rlx\hbox{\rm I\kern-.18em P}}
\def\IQ{\rlx\hbox{\,$\inbar\kern-.3em{\rm Q}$}}
\def\IF{\rlx\hbox{\rm I\kern-.18em F}}
\def\IG{\rlx\hbox{\,$\inbar\kern-.3em{\rm G}$}}
\def\IH{\rlx\hbox{\rm I\kern-.18em H}}
\def\II{\rlx\hbox{\rm I\kern-.18em I}}
\def\IK{\rlx\hbox{\rm I\kern-.18em K}}
\def\IL{\rlx\hbox{\rm I\kern-.18em L}}
\def\one{\hbox{{1}\kern-.25em\hbox{l}}}
\def\0#1{\relax\ifmmode\mathaccent"7017{#1}%
B        \else\accent23#1\relax\fi}

\def\PRL#1#2#3{{\sl Phys. Rev. Lett.} {\bf#1} (#2) #3}

\def\PRD#1#2#3{{\sl Phys. Rev.} {\bf D#1} (#2) #3}
\def\PRA#1#2#3{{\sl Phys. Rev.} {\bf A#1} (#2) #3}
\def\PRE#1#2#3{{\sl Phys. Rev.} {\bf E#1} (#2) #3}

\def\PLA#1#2#3{{\sl Phys. Lett.} {\bf #1A} (#2) #3}

\def\JMP#1#2#3{{\sl J. Math. Phys.} {\bf #1} (#2) #3}

\def\RMP#1#2#3{{\sl Rev. Mod. Phys.} {\bf #1} (#2) #3}
\def\PR#1#2#3{{\sl Phys. Reports} {\bf #1} (#2) #3}

\def\JPAMG#1#2#3{{\sl J. Physics A: Math. Gen.} {\bf A#1} (#2) #3}

\def\JPA#1#2#3{{\sl J. Physics} {\bf A#1} (#2) #3}
\def\JPA#1#2#3{{\sl J. Physics A: Math. Theor.} {\bf A#1} (#2) #3}

\def\PHSD#1#2#3{{\sl Physica} {\bf D#1} (#2) #3}

\def\JHEP#1#2#3{{\sl JHEP} {\bf #1} (#2) #3}

\def\JCP#1#2#3{{\sl Journal of Computational Physics} {\bf #1} (#2) #3}
\def\SAM#1#2#3{{\sl Stud. Appl. Math.} {\bf #1} (#2) #3}
\def\MAA#1#2#3{{\sl Methods and  Applications of Analysis} {\bf #1} (#2) #3}
\def\Nonl#1#2#3{{\sl Nonlinearity} {\bf #1} (#2) #3}
\def\OL#1#2#3{{\sl Optics Letters} {\bf #1} (#2) #3}

\begin{document}

\begin{titlepage}

\vspace{.2in}
\begin{center}
{\large\bf Spatial shifts of colliding dark solitons in deformed non-linear Schr\"{o}dinger models} 
\end{center}

\vspace{.5in}

\begin{center}

Harold Blas$^{a}$  and Marcos Zambrano$^{b}$

\par \vskip .2in \noindent

$^{a}$ Instituto de F\'{\i}sica\\Universidade Federal de Mato Grosso\\
Cidade Universit\'aria, Cep 78060-900, Cuiab\'a - MT - Brazil\\ 

$^{b}$ Instituto de Ci\^encias Matem\'aticas e de Computa\c c\~ao; ICMC/USP\\ 
Universidade de S\~ao Paulo\\
Caixa Postal 668, CEP 13560-970, S\~ao Carlos-SP, Brazil\\
\normalsize
\end{center}

\vspace{.3in}

\begin{abstract}
\vspace{.3in}

We derive a closed expression for the spatial shift experienced by a black soliton colliding with a shallow dark soliton in the context of deformed non-linear Schr\"{o}dinger models. A perturbative scheme is developed based on the expansion parameter $1/(\g v)<<1$, where $v$ is the velocity of the incoming shallow dark soliton, $\g \equiv 1/\sqrt{1-v^2/v_s^2}$\, and  $v_s$ is the Bogoliubov sound speed, therefore it is not restricted to small deformations of the integrable NLS model. As applications of our formalism we consider the integrable NLS model and  the non-integrable cubic-quintic NLS model with non-vanishing boundary conditions. Extensive numerical simulations are performed in order to verify our results. A variant of the analysis for gray-gray soliton collision is discussed regarding a fast broad soliton and a slow thin soliton.
\end{abstract}

\end{titlepage}
\section{Introduction}
 
The mathematical theory of soliton collisions in the integrable models is a well developed subject. In particular, the solitons emerge with their properties completely unchanged after collision between them, except for relevant phase shifts. However, certain non-linear field theory models possess solitary wave solutions and it is difficult to know {\sl a priori} if they are in fact true solitons with analytical multisoliton solutions. Regarding this issue the so-called quasi-integrability concept has recently been put forward related to the deformations of the integrable sine-Gordon (SG) and non-linear Schr\"{o}dinger (NLS) models \cite{jhep1}. According to this new concept certain deformed integrable field theory models share some properties with their truly integrable counterparts for some configurations, for example they possess an infinite number of quasi-conserved charges. For the case of two solitary wave collision those charges are asymptotically conserved in the scattering process, although the charges vary in time their values in the far past and the far future remain the same. The both numerical and analytical tools were used to define and describe the concept of quasi-integrability.   

Our aim is to predict the results of solitary wave collisions in  deformed NLS models with non-vanishing boundary conditions. Such theories appear in diverse areas of non-linear science, such as condensed matter physics, plasma physics  and,  in particular, in the study of Bose-Einstein condensates. We are also motivated by the above mentioned quasi-integrability property in deformed NLS models with non-vanishing boundary conditions in which sufficiently fast solitary waves effectively pass through each other. 

In the context of analytical calculations some results have been obtained only for the cases of small perturbations of the NLS model \cite{keener, malomed, ablowitz}. For nearly integrable models it has been implemented a perturbation theory allied to the the inverse scattering transform (IST) method \cite{malomed1}. These methods have been applied mainly to the NLS solitons that decay at infinity, i.e. the bright solitons. They encompass the above mentioned perturbation theory based on IST and a direct perturbation theory \cite{herman}. On the other hand, the dark solitons are intensity dips on a continuous wave background with a phase change across their intensity minimum. However, the non-vanishing boundary of dark solitons, which may change when the deformations are present, introduces serious complications when applying the perturbative techniques developed for bright solitons. In the last years, various improvements  have been made to the calculations and methods \cite{chen, lashkin, ao} based on the so-called complete set of squared Jost solution (eigenvectors of the linearized NLS operator) in order to apply to the non-vanishing boundary conditions and dark soliton perturbations. The
implementation of direct methods, however, consider small perturbations around the NLS model \cite{ablowitz, yu}. Moreover, in \cite{ablowitz} it has been argued that the squared Jost functions associated with the dark soliton might be an insufficient basis, so rendering problematic 
 the issue of a complete basis for the solution space of the deformed NLS equation. 

Here we propose a method to study shallow solitary wave collisions in the context of a perturbation theory performed order by order in an expansion parameter related to the both velocity of the incoming soliton and the Bogoliubov sound speed. The expansion parameter has a kinematic origin and does not appear explicitly in the Lagrangian of the model. In this way our series expansion is not restricted to small deformations around the integrable NLS model. We will assume that at zeroth order in this parameter the colliding solitary waves effectively pass through each other and provide the leading order contribution to the spatial shift experienced by a stationary soliton  after colliding with an incoming soliton. Similar perturbative framework has recently been considered in the study of scattering of ultra-relativistic kink-type solitary waves of the deformed SG model \cite{amin1, amin2}. We present the relevant developments in the context of a non-Lorentz invariant field theory associated to the NLS complex field and show that the absolute value of the spatial shift  experienced by a black soliton after collision with an shallow gray soliton decreases as the  velocity increases, in this way resembling the behavior of ultra-relativistic kink-(anti)kink collisions.

It has recently been shown that the perturbed dark soliton of the NLS model comprises an inner region, containing the core of the soliton and a moving shelf, and an outer region which matches to the boundary at infinity \cite{assanto, ablowitz}. The main features of those results relevant to our discussions here are the appearance of a shelf of finite length around the both sides of the soliton the edge of which propagates with speed determined by the background intensity, and that the difference between the amplitude of the shelf and the outer background is very small compared with the amplitude of the background intensity.

The main characteristics  of our perturbative scheme 
are the following. We consider a linear superposition of two solitary wave solutions as a good background solution during collision, such that the interaction of the colliding solitons occurs significantly in the overlapping region which is a region contained in the inner region of the perturbed stationary solitary wave. The continuous wave background will be subtracted on top of the linear superposition in order to maintain the non vanishing boundary condition at infinity. So, we are interested in investigating the effects of short-range interactions between the dark solitons and the long range contributions arising from the shelves mentioned above will be ignored in the integrals employed to compute the spatial shift. In the rest frame of the stationary solitary wave where the other approaches with velocity $|v| \rightarrow v_s$, the  space-time integration of the overlapping region becomes proportional to $\frac{1}{v \g}$, where $\g = 1/\sqrt{1-(v/v_s)^2}>>1$,  $v_s$ being the Bogoliubov sound speed. This property is systematically used to provide a perturbative framework in order to  calculate the spatial shift experienced by the stationary soliton, using the small value of $(\frac{1}{v \g})$ as the parameter of our perturbation expansion. The perturbation parameter $(\frac{1}{v \g})$ arises after space-time and suitable field space  transformations are performed simultaneously. In this regard the implementation of the perturbative analyses of the non-relativistic deformed NLS model is  slightly different from the relativistic deformed SG model in which an analogous factor appears directly from a coordinate transformation to the rest frame of the static soliton \cite{amin1}.   

The paper is organized as follows. In section \ref{sec2} we discuss a deformed NLS model with  non-vanishing boundary conditions and the relevant equations leading to dark soliton solutions. In section \ref{sec3} we develop a perturbative scheme in order to compute the spatial shift experienced by a black soliton after colliding with a gray soliton. In section \ref{sec4} we present some examples of application of our formalism. The integrable NLS model is considered in some detail in order to check our perturbative scheme. Afterwards, the collision of dark solitons of the cubic-quintic NLS model is studied as an example of a non-integrable theory. In section \ref{sec5} we develop numerical simulations for the both models considered in section \ref{sec4}. In section  \ref{sec6} we discuss the main results and possible future directions of research. Finally, in the appendix \ref{dark} the properties of dark soliton interaction in the integrable NLS model are presented, and  in the appendix \ref{cq1} some properties of dark solitons of the cubic-quintic NLS model are discussed 
        
\section{Deformed NLS models and solitary waves}
\label{sec2}

Various generalizations and deformations of the integrable NLS model arise in physical applications. We will consider the deformed NLS models of the type
\br
\label{nlsd}
i \frac{\pa}{\pa t} \psi(x,t) + \frac{1}{2} \frac{\pa^2}{\pa x^2} \psi(x,t) -  \frac{\pa f[|\psi(x,t)|^2]}{\pa |\psi(x,t)|^2} \psi(x,t) =  0,\er
where $f: \IR_{+} \rightarrow \IR $. The model (\ref{nlsd}) allows dark soliton type solutions in analytical form for certain functions $f[I]$; for example, the integrable  NLS case corresponds to $f[I] = \b I^2/2$ (see Appendix \ref{dark}) and  the non-integrable cubic-quintic NLS (CQNLS) defined by $f[I] = \b I^2/2-\a I^3/6$ \cite{crosta} (see Appendix \ref{cq1}). Among the models with saturable non-linearities \cite{kivshar}, the case $f[I] = \frac{1}{2} \rho_s (I+\frac{\rho_s^2}{I+ \rho_s})$ also exhibits analytical dark solitons \cite{kroli}. The qualitative properties of traveling waves of the NLS model for general non-linearities and non-vanishing boundary conditions have been studied in \cite{chiron}. The formalism developed here to study soliton collisions will require the deformed  NLS model to possess dark-soliton type solutions obtained either analytically or numerically, as well as the condition that the linear superposition of two such solitons provides a good background solution during their collision.  

We will consider translationally invariant solutions of the form 
\br
\label{dark00}
\psi(x,t) = \phi(z) \mbox{exp}[i\theta(z)+ i w t],\,\,\,\,z= x- v t,
\er
satisfying the following non-vanishing boundary condition ({\sl nvbc})
\br
\label{bc}
|\phi(z \rightarrow \pm \infty)| = |\psi_0|,\,\,\,\,\, \pa_{z} \phi_{(z \rightarrow \pm \infty)} =  0, \,\,\,\,\,\pa_{z} \theta_{(z \rightarrow \pm \infty)} = 0.   
\er   

The {\sl nvbc} (\ref{bc}) is suitable for  wave solutions in the form of localized `dark' pulses created on the continuous wave background ({\sl cwb}) at rest  
\br
\label{cwb}
\psi_0(x,t) = |\psi_0| e^{i \( w t + \d\)},
\er
where $ w = - f'[|\psi_0|^2]$ (here and in the upcoming sections $f'[y] \equiv \frac{d f[y]}{d y}$) is the phase shift experienced by the {\sl cwb} with intensity $|\psi_0|^2$, in the non-linear medium with intensity $|\psi|^2$ and varying according to the law $-  \frac{\pa f[|\psi|^2]}{\pa |\psi|^2}$.  Moreover, we will consider a family of systems such that small oscillations around $|\psi|=|\psi_0|$ of (\ref{nlsd}) defines a hyperbolic system with sound speed
\br
v_s = |\psi_0| \sqrt{f''[|\psi_0|^2]},\,\,\,\,\,f''[|\psi_0|^2]> 0.
\er

\subsection{Solitary waves}
\label{solwaves}

Among the solutions of type (\ref{dark00}) let us assume some dark solitons as follow 
\br
\label{para1}
\psi_1(x,t) &=& \phi_{1}(x) e^{i w t}\, e^{i\zeta},\\
\psi_2(x,t) &=& \phi_{2}(z) e^{i \theta(z)} e^{i w t}\, ,\,\,\,\,\,z = x - vt,\label{para2}
\er
where $\psi_1$ will be the stationary `black' soliton and $\psi_2$ will be  the incoming `gray' soliton. $\phi_1,\,\phi_2,\theta$ are smooth functions of their arguments such that  $\phi_1: \IR \rightarrow [-|\psi_0|,|\psi_0|],\,\,\phi_2: \IR \rightarrow [q_0, |\psi_0|] \subset \IR_{+},\,\,\theta: \IR \rightarrow [\theta^{-},\theta^{+}]$. Moreover, they converge exponentially fast to the non-vanishing boundary conditions: $\phi_{1}(\pm \infty) \rightarrow \pm |\psi_0|$,\,\,$\phi_{2}(\pm \infty) \rightarrow  |\psi_0|$,\,\, $\theta(\pm \infty) \rightarrow  \theta^{\pm}$. The constant phase $\zeta$ in (\ref{para1})  has been introduced for later purpose, and the frequency $w$ has been assumed to be the same as the one of the  {\sl cwb} in (\ref{cwb}). 

One can write an equation for $\phi_1$ by substituting the Ansatz for $\psi_1$ (\ref{para1}) into (\ref{nlsd})
\br
\label{phi1}
\phi''_1 - 2 (w + f'[\phi_1^2] ) \phi_1 = 0. 
\er
From the equation (\ref{phi1}) one can get the  relationships
\br\nonumber
\frac{d \phi_1}{d x} &=& \pm \sqrt{2} \Big[\int_{|\psi_0|^2}^{\phi_1^2} \( f'[I]- f'[|\psi_0|^2]\) dI\Big]^{1/2}\\
&\equiv & \pm F[\phi_1] \label{ff1}
\er
 
The corresponding solution for the NLS model is a black soliton $\phi_1$, the solution (\ref{dark111}) with $v=0$,  which is an antisymmetric function of space and develops an abrupt $\pi$ phase shift and zero intensity at its center. The $\mp$ signs in (\ref{ff1}) correspond to the regions $x\in[-\infty, x_0]$ and $x \in [x_0, +\infty]$, respectively, where $x_0$ represents the zero intensity position of the black soliton. The relevant black soliton  $\phi_1$ of the CQNLS model is presented in (\ref{bl1}) which presents quite similar properties.  

Substituting the Ansatz for $\psi_2$ (\ref{para2}) into (\ref{nlsd}) one gets the following equations
\br
\label{phi2}
\phi''_2 - 2 (w + f'[\phi_2^2] ) \phi_2 + v^2\(\phi_2 -\frac{|\psi_0|^4}{\phi_2^3} \) &=&0\\
\frac{d}{dz}\Big\{\phi^2_2 [\theta'(z) - v] \Big\} &=& 0\label{theta},
\er  
where $w = - f'[|\psi_0|^2] $. Integrating the last equations  one gets
\br
 \frac{d \phi_2}{d z} &=& \pm \sqrt{2} \Big[\int_{|\psi_0|^2}^{\phi_2^2} \( f'[I]-f'[|\psi_0|^2]\) dI - \frac{v^2}{2} \frac{(\phi_2^2-|\psi_0|^2)^2}{\phi_2^2} \Big]^{1/2}\nonumber\\
&\equiv & \pm G[\phi_2] \label{ff2}\\\nonumber
\frac{d \theta}{d z} &=& v \(1-\frac{|\psi_0|^2}{\phi_2^2}\)\\
&\equiv & H[\phi_2] \label{ff3}\\ \nonumber
\theta(x,t) &=&\pm  \int_{|\psi_0|}^{\phi_2} \frac{H[y]}{G[y]} dy\\
               &\equiv &  \pm \Theta[\phi_2]  \label{id1}
\er

In the NLS case one has the solution (\ref{dark111}) such that $\phi_2 = |\psi(x,t)|$ and $\theta$ being the space dependent phase of $\psi(x,t)$.  Whereas, the gray soliton  $\phi_1$ of the CQNLS model is presented in (\ref{sl1}) and the space dependent phase in (\ref{th1}).  Notice that in the both models the $\mp$ signs in (\ref{ff2}) correspond respectively to the regions $z\in[-\infty, z_0]$ and $z \in [z_0, +\infty]$  of the dark soliton, where $z_0$ is the coordinate in which the soliton amplitude approaches its minimum intensity (dip). 

The analytical expression for the two-dark soliton solution of the integrable NLS model is presented in (\ref{2dark}) which allows the computation of the exact value of the spatial shift developed after collision of black and gray solitons (\ref{shiftnls1}). In Fig. 1 we have plotted a collision process of these solitons.

The above relationships (\ref{ff1}) and (\ref{ff2})-(\ref{id1}) will be used below in our study of the soliton collisions in the framework of a perturbative scheme and in the Appendix \ref{cq1} in order to find the CQNLS dark solitons.

\section{Perturbative expansion for black and gray soliton collision}
\label{sec3}

In order to study dark soliton collisions we will assume that the system (\ref{nlsd}) possesses some solitary waves such that their linear combination provides a good background solution during collision. For variety of functionals $f[I]$ and in the relevant ones used in the examples below this property is verified. Consider a solution before, during and after the collision of two-dark solitons
\br \Psi(x,t) \equiv \psi_{1}(x,t) + \psi_{2}(x,t) + h(x,t) - \psi_0(x,t).
 \label{total}\er

Here $\psi_1(x,t)$ represents the stationary `black' soliton (\ref{para1}) and $\psi_2(x,t)$ the moving `gray' soliton (\ref{para2}). The function $h(x,t)$ describes the perturbation effects around the core of the stationary black soliton generated by the collision. It has been included the continuous wave background (\ref{cwb}) in order to subtract the {\sl nvbc} reference term  $\psi_0(x,t)$ on top of the superposition, since the functional $f[I]$  associated to the potential of the system is not periodic. The constant phase parameters $\zeta$ and $\d$ associated to the black soliton (\ref{para1}) and the {\sl cwb} (\ref{cwb}), respectively, will play below an important role in order to compute the spatial shift. The Fig. 2 shows a gray-black soliton system of the CQNLS model before, during and after collision. Notice the successive deformations experienced by the static soliton during the entire process, in particular the Fig. 2.c. shows  the deformation, described by the function $h(x,t)$ above, of the core of the static soliton during the nearest position of the both solitons.        

As mentioned above a dark soliton under perturbation develops a shelf the edge of which propagates out on the both sides of the soliton. In order to study the radiation shed by an evolving dark soliton one can linearize the NLS equation (\ref{nlsd}) about the ({\sl cwb}) as $ \psi_{a}(x,t) \approx |\psi_0| e^{i w t } + |\psi_{0a}| e^{iw t }$. So, the moving shelves contributions associated to each soliton can be incorporated as $\psi_{a}(x,t) + |\psi_{0a}| e^{iw t }$ ($a=1,2$); however it has been shown that $|\psi_{0a}| << |\psi_{0}|$  \cite{assanto, ablowitz}, therefore we will neglect them in our computations of the spatial shift at first order in $1/(v\g)$.    

The equation (\ref{nlsd}) can be linearized in order to find the equation for $h(x,t)$. So, one has 
\br i\partial_{t}h + \frac{1}{2}\partial_{x}^{2}h - W_1(x,t) h - W_2^1(x,t) h - W_2^2(x,t) \bar{h} &=& S(x,t)\nonumber \\
&+& \D W_1(x,t) h + \D W_2^1(x,t) h + \D W_2^2(x,t) \bar{h} \label{eqh}
\er
where
\br
W_1(x,t) &\equiv & f'[|\psi_{1}|^{2}],\\
W_2^1(x,t)  &\equiv & f''[|\psi_{1}|^{2}] |\psi_{1}|^{2},\\
W_2^2(x,t) &\equiv & f''[|\psi_{1}|^{2}] \psi_{1}^{2},\\
\nonumber
S(x,t) &\equiv & (f'[|\Psi_{0}|^{2}] - f'[|\psi_{1}|^{2}])\psi_{1} + (f'[|\Psi_{0}|^{2}] - f'[\psi_{2}])\psi_{2} \\ 
&& +(f'[|\psi_{0}|^{2}] - f'[|\Psi_{0}|^{2}])\psi_{0}  \label{source}\\
\label{d1} \D W_1(x,t) & \equiv &  f'[|\Psi_{0}|^{2}] - f'[|\psi_{1}|^{2}],\\
\label{d2} \D W_2^1(x,t) & \equiv & f''[|\Psi_{0}|^{2}]|\Psi_{0}|^{2} - f''[|\psi_{1}|^{2}]|\psi_{1}|^2, \\
\label{d3} \D W_2^2(x,t)  & \equiv & f''[|\Psi_{0}|^{2}](\Psi_{0})^2 - f''[|\psi_{1}|^{2}]\psi_{1}^{2}, \\
\Psi_0(x,t) & \equiv & \psi_{1}(x,t) + \psi_{2}(x,t) - \psi_0(x,t).\label{d44}
\er
The expressions $W_1, W_{2}^1$ and $W_{2}^2$ in the l.h.s. of (\ref{eqh}) are the potential terms for the perturbation around an isolated stationary dark soliton $\psi_1$; and  $\D W_{1}, \D W_{2}^1$ and $\D W_{2}^2$ are their relevant changes due to the incoming dark soliton $\psi_2$. We refer to $S(x,t)$ as the external source, whose real and imaginary parts behave as step-like functions around the regions occupied by the solitons, becoming a single step-like function when the two solitons overlap as it was shown in the Fig. 3 for the collision of gray and black solitons of the CQNLS model. Notice that in the absence of the gray soliton one can make $h=0,\, \psi_2 \rightarrow \psi_0$, then $S, \D W_1, \D W_2^1$ and $\D W_2^2$ vanish identically, whereas they are non vanishing in the regions occupied by the gray soliton $\psi_2$. Considering the incoming dark soliton as the source of perturbation one notices that $h, S,\D W_1, \D W_2^1 , \D W_2^2$ are non vanishing when the two dark solitons overlap.
 
There is another  equation for $\bar{h}$, where the bar stands for complex conjugation, which can be obtained by taking the complex conjugate of (\ref{eqh}).   

\begin{figure}
\centering
\label{fig1}
\includegraphics[width=10cm,scale=4, angle=0, height=6cm]{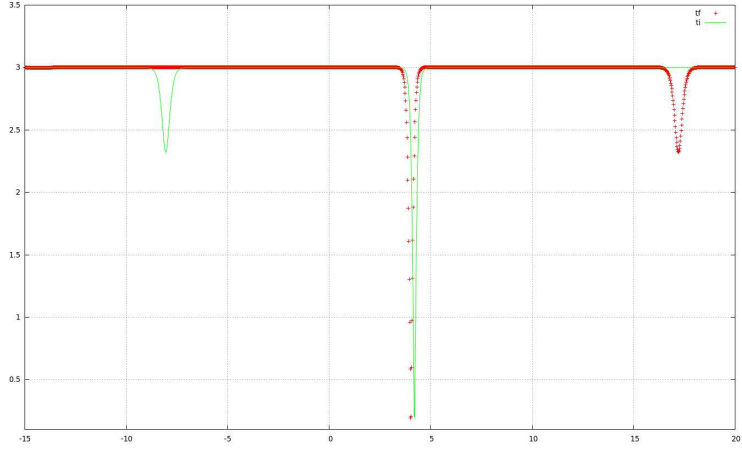} 
\parbox{5in}{\caption {(color online) The initial condition (\ref{dark2})-(\ref{para11}) for a dark-black soliton system of the NLS model (\ref{nls2}) is plotted for $\b=4,\,|\psi_0|=3,\,\zeta =1.15026$ as a continuous line (green). The gray soliton initially located at $x_2=-8$ travels to the right with velocity $v = 4.64$. The black soliton is located at $x_1=4.2$. The final configuration after collision is the dotted line (red). Note the spatial shift $\D x \approx -0.124 $ experienced by the black soliton.}}
\end{figure}

\begin{figure}
\label{fig2}
\includegraphics[width=8cm,scale=4, angle=0, height=5cm]{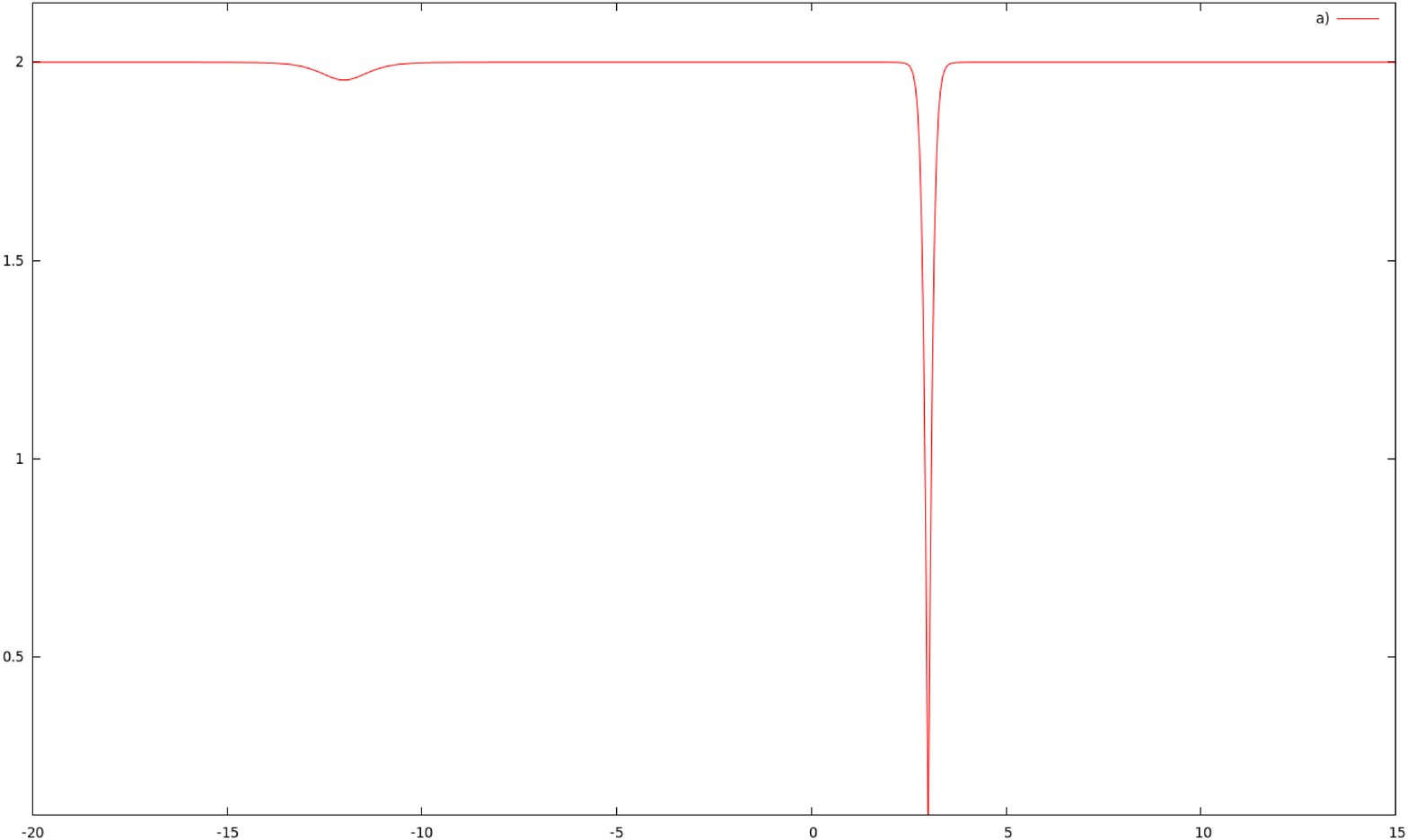}
\includegraphics[width=8cm,scale=4, angle=0, height=5cm]{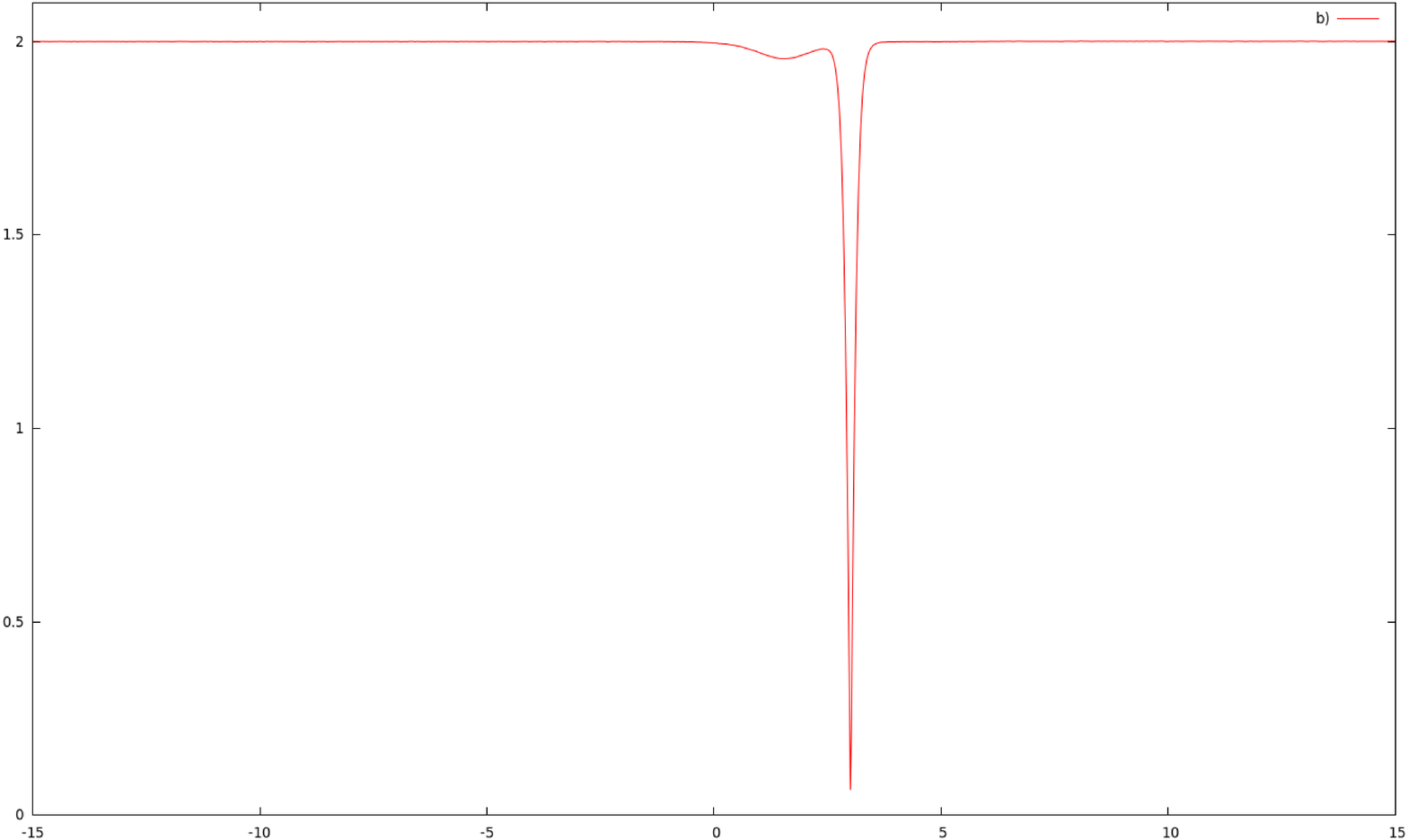}
\includegraphics[width=8cm,scale=4, angle=0, height=5cm]{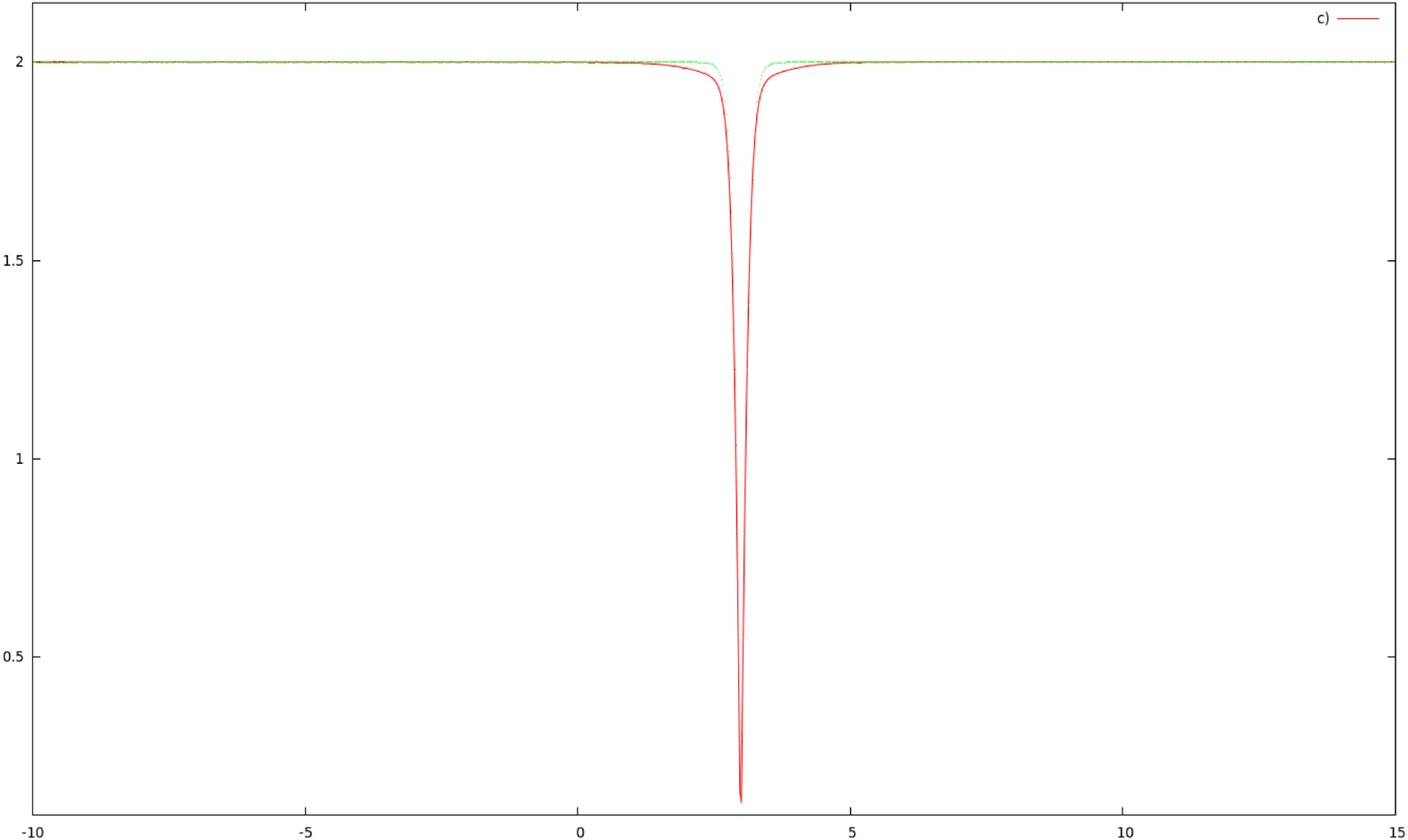}
\includegraphics[width=8cm,scale=4, angle=0, height=5cm]{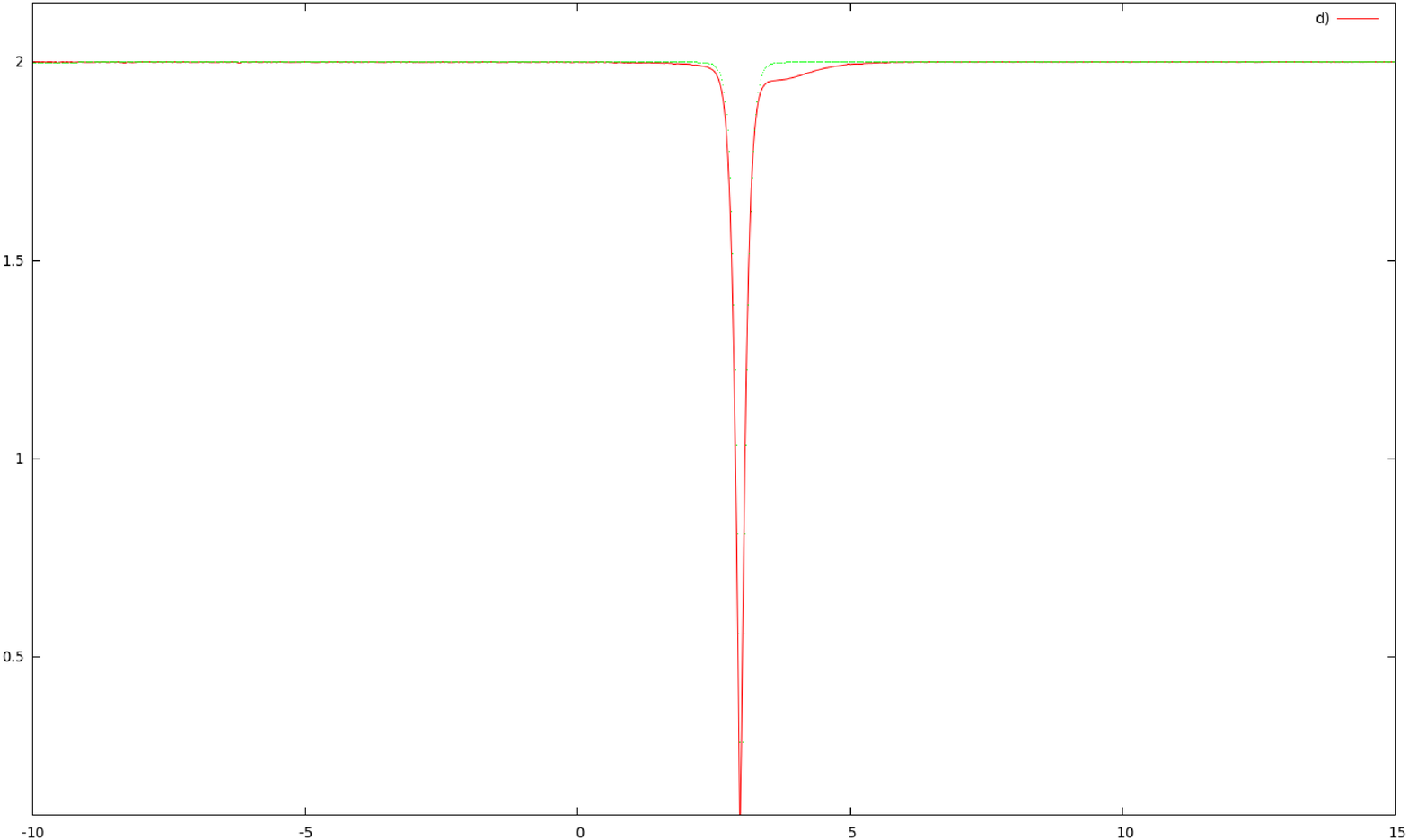}
\includegraphics[width=8cm,scale=4, angle=0, height=5cm]{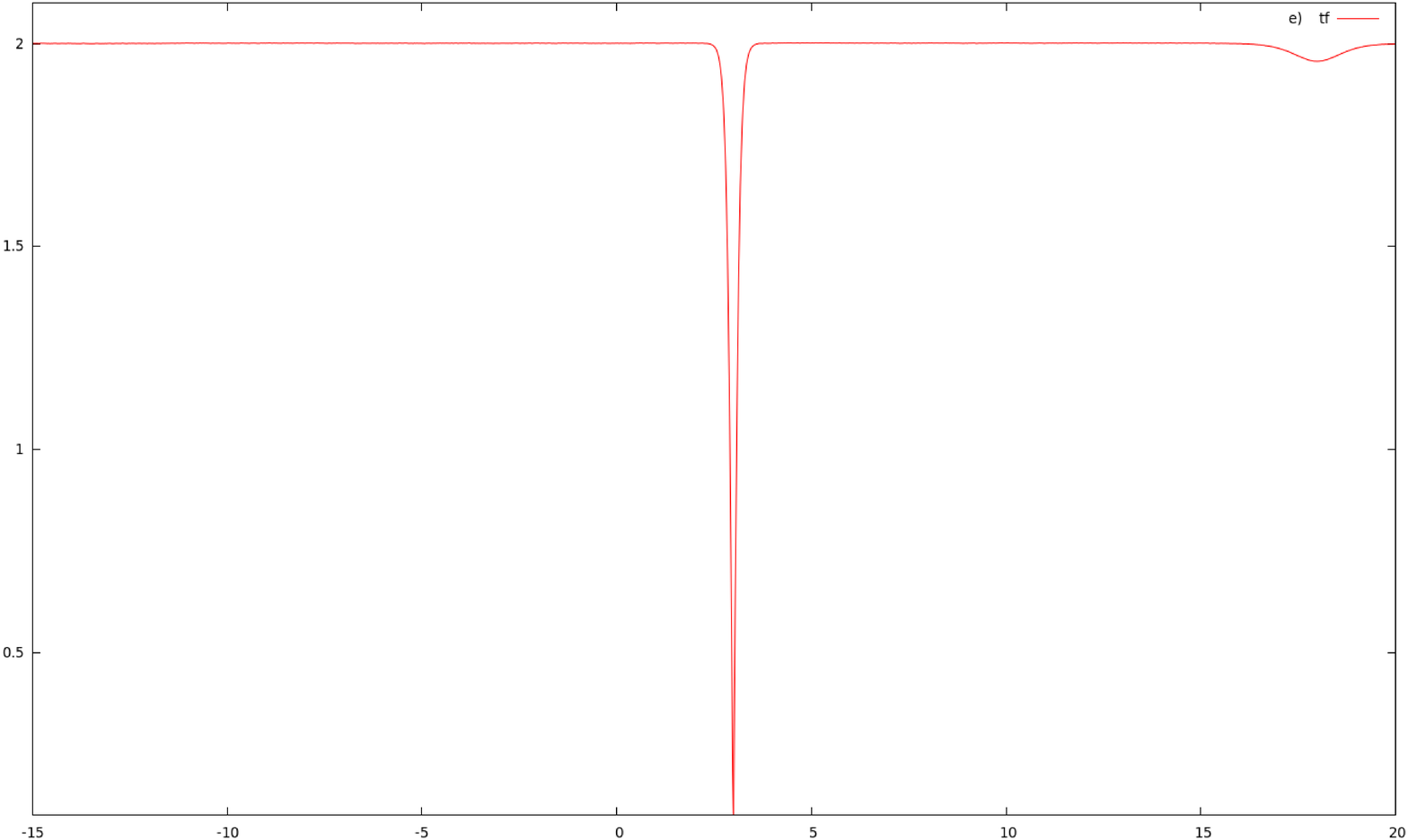}
\includegraphics[width=8cm,scale=4, angle=0, height=5cm]{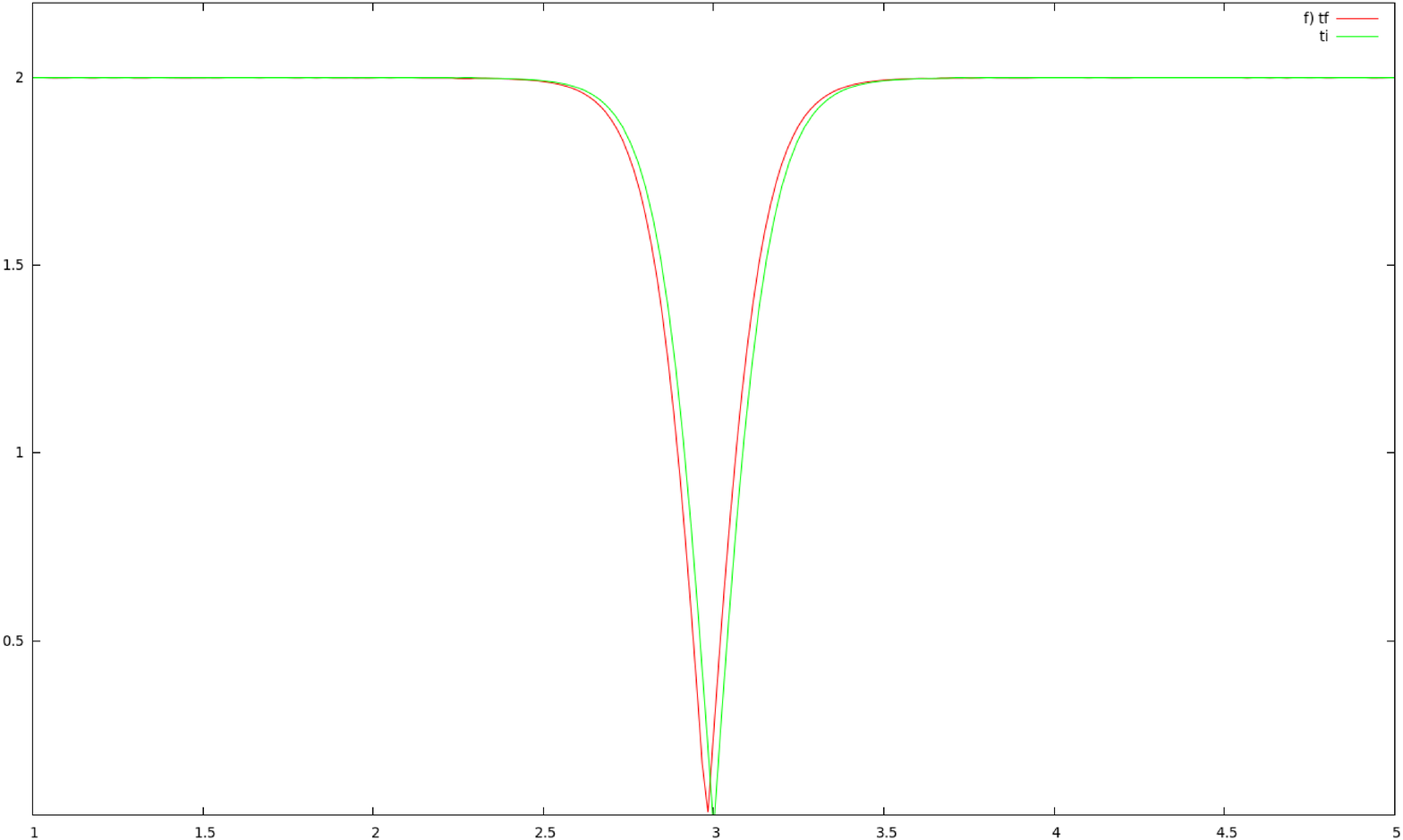}

\parbox{6.5in}{\caption{(color online) a) Before collision. The initial red profiles of the CQNLS dark solitons. The left soliton is the gray soliton and the right one is the static black soliton. b) The gray soliton is approaching the black soliton from the left. c) The nearest position of the both solitons. For comparison the initial profile of the black soliton is plotted as dotted green line. d) The gray soliton pulse is starting to emerge to the right from the position of nearest collision. The initial black soliton is the green curve. e) After collision. The final profile of the dark solitons. The right soliton is the gray soliton and the left one is the static black soliton. f) The spatial shift of the black soliton. The final soliton (red curve left) has a position shifted  with respect to the initial black soliton position (green curve right). The parameters are $\a = 0.6,\,\b=12.0,\,|\psi_0|=2.0,v=6.07,\,v_s=6.19677$.}}
\end{figure}

\begin{figure}
\centering
\label{fig3}
\includegraphics[width=10cm,scale=4, angle=0, height=5.5cm]{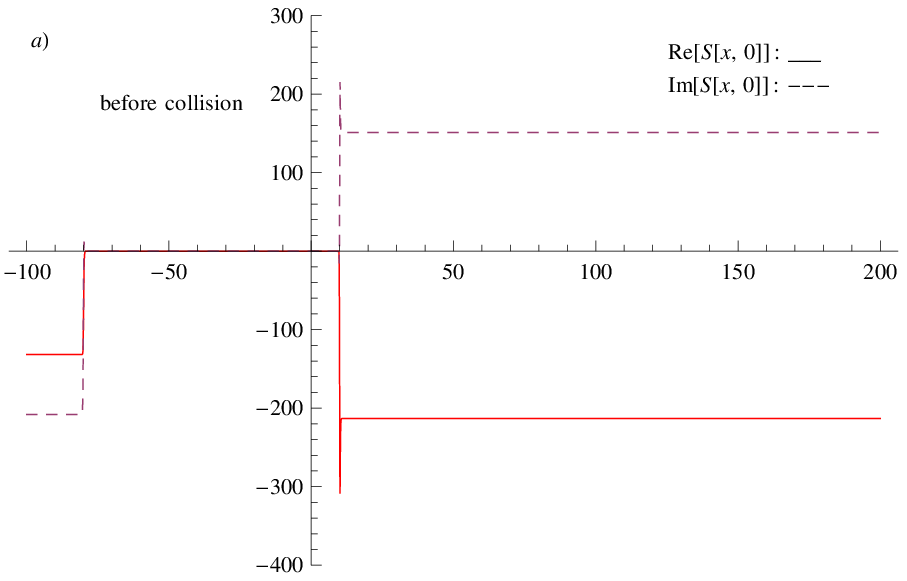}
\includegraphics[width=10cm,scale=4, angle=0, height=5.5cm]{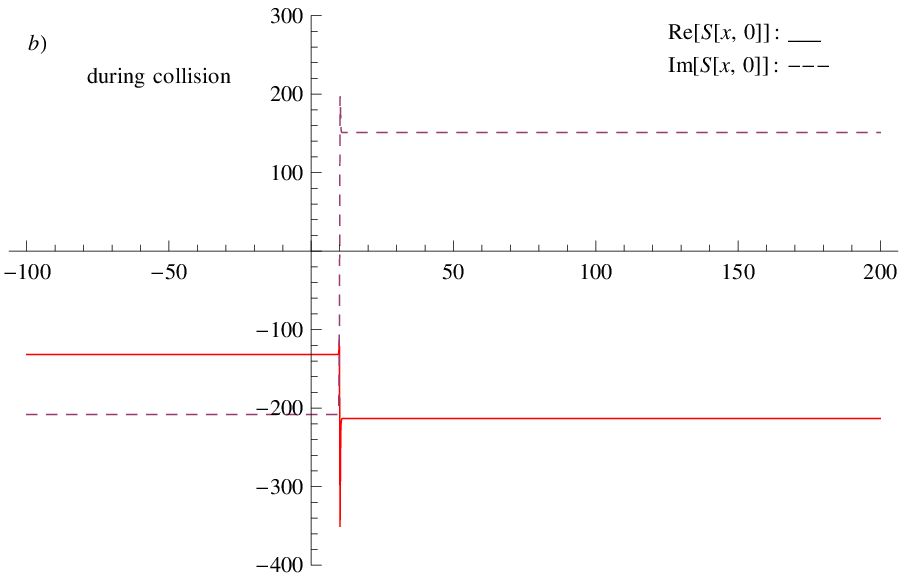}
\includegraphics[width=10cm,scale=4, angle=0, height=5.5cm]{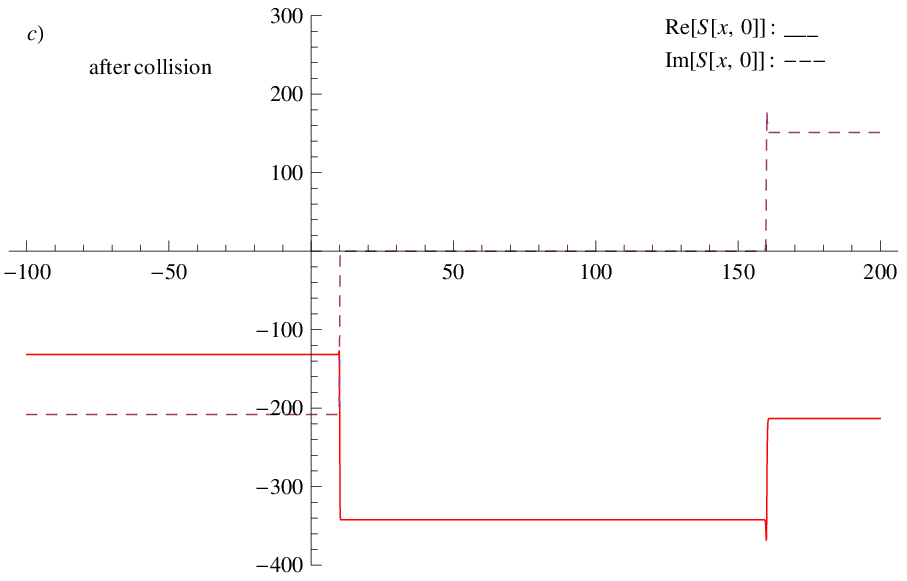}
\parbox{6in}{\caption {(color online) The real (continuous curves) and imaginary (dashed curves) parts of the source $S(x,t)$ generated by the incoming gray soliton through its interaction with the black  soliton of the CQNLS model. The initial profile (\ref{dark22})-(\ref{para22c}) has been used for the gray-black soliton system. The gray soliton approaches the black soliton from the negative direction of the coordinate $x$. a) The initial profile of the source. The left step-like functions are around the gray soliton position and the right ones are around the static black soliton. b) The source $S$ during the collision and the nearest position of the both solitons. c) The source profile after the collision.}}
\end{figure}

In order to find solutions of the eq. (\ref{eqh}) we expand $h(x,t)$ as 
\br
\label{eigen}
h(x,t) = \sum_{\l} g_{\l}(t) u_{\l}(x)
\er  
where $\{ u_{\l}(x)\}$ form an orthonormal basis and solve the eigenvalue equation in matrix notation
\br
\label{bdg}
{\cal H}_{x} \left[\begin{array}{c} u_{\l} \\ \bar{u}_{\l}\end{array}\right] = \l \left[\begin{array}{c} u_{\l} \\ \bar{u}_{\l}\end{array}\right]  
\er

\br 
{\cal H}_{x} \equiv \left[\begin{array}{cc}
-\frac{1}{2}\partial_{x}^{2} + f'[|\psi_1|^2] + f''[|\psi_1|^2] |\psi_1|^2 & f''[|\psi_1|^2] (\psi_1)^2  \\
 f''[|\psi_1|^2] (\bar{\psi}_1)^2 & -\frac{1}{2}\partial_{x}^{2} + f'[|\psi_1|^2] + f''[|\psi_1|^2] |\psi_1|^2 \end{array} \right]
\er

A remarkable fact is that one can find the first zero modes ($\l = 0 $)
 \br\label{zeromodes} 
u_{0}^1(x) = e^{i \frac{\pi}{2}} \psi_1,\,\,\,\, u_{0}^2(x) = \pa_x \psi_1,\er
 where $\psi_1$ solves the stationary version of the deformed NLS equation (\ref{nlsd}) for any $f[I]$.  These zero modes are associated to the phase and translational symmetries of the model, respectively. Regarding this issue, let us mention that the linearized eigenvalue equation around a dark soliton solution in the integrable NLS model possesses four linearly independent zero modes $u_{0}^j(x,t), j=1,...,4$ and their exact analytical expressions have been presented in \cite{sykes}. Those zero-modes are associated to the following four fundamental symmetries of the NLS model: phase, translational, Galilean and scale invariance, respectively.

\subsection{spatial shift and a perturbative expansion}

Next, we make use of the linearized equation (\ref{eqh}) in order compute  the spatial shift experienced by the black soliton after collision with a gray soliton. It is considered that the shallow gray soliton is approaching the static black soliton from the negative $x$ direction as in Figs.1 and 2. Let us expand the stationary black soliton as
\br
\label{taylor}
\psi_{1}( x+\D x(t),t)=  \psi_{1}(x,t)+ \pa_x \psi_{1}( x,t) \D x(t)+...
\er
Then in order to extract the spatial shift we take
\br
\label{exp}
h(x,t) = \pa_x \psi_{1}( x,t) \D x(t)+...
\er

The function $\pa_x \psi_{1}( x,t)$ above is precisely the zero mode $u_{0}^{2}(x)$ (\ref{zeromodes}) associated to the space translation symmetry of the deformed NLS model.  Comparing (\ref{exp}) with the eigenmode expansion  as in (\ref{eigen}) one can see that the ellipses represent the expansion terms in the eigenmodes which are orthogonal to $\pa_x \psi_{1}( x) $. Regarding the velocity change of the black soliton $\D V$ one can consider a term like $h(x,t) = u_{0}^{3}(x) \D V +...$, where $u_{0}^{3}(x)$ is the zero mode associated to Galilean symmetry of the deformed NLS model. However, as we will show below the spatial shift does not present a time dependence at leading order of perturbation theory, so the velocity change $\D V$ must be zero at this order. This fact is also verified through our numerical simulations for some concrete examples. We are mainly interested in the effects of the zero mode associated to space translation and postpone for a future work the treatment of the remaining zero modes and higher order effects.     

Since $\D W_1, \D W_2^1\,$ and  $\D W_2^2$  are multiplied by $h$ in (\ref{eqh}) their effect are small compared to the source term $S$, so substituting (\ref{exp}) into (\ref{eqh}) and considering only the source term $S$ one can get 
\br
\label{delta}
i   \pa_{x} \psi_1(x,t) \frac{ d \D x (t)}{dt}  = S(x,t)  
\er 

Projecting out the last relation and its complex conjugate onto the zero modes associated to space translation symmetry $\pa_{x} \psi_1$ and  $\pa_{x} \bar{\psi}_1$, respectively, and integrating with respect to  $ x $ one gets 

\br
\label{dxt1}
\frac{ d \D x (t)}{dt} & = & \( \frac{1}{{\cal D}(t) } \) i \int_{-\infty}^{+\infty} \, dx \,  \Big[  \pa_{x} \psi_1 \bar{S}(x,t)- \pa_{x} \bar{\psi}_1 S(x,t)\Big]\\
 {\cal D} (t)& \equiv & 2 \int_{-\infty}^{+\infty} dx \,  \pa_{x} \bar{\psi}_1  \pa_{x} \psi_1. \label{dxt2}
\er 

Consider the initial condition $\D x (t\rightarrow -\infty) = 0$, then integrating (\ref{dxt1}) w.r.t.  $t$ one has
\br
\label{d0}
\D x (t) &=& \frac{{\cal N}(t)}{{\cal D}(t)}\\
{\cal N}(t) & \equiv &  i \int_{-\infty}^{t} d\tau \int_{-\infty}^{+\infty} \, dx \,  \Big[  \pa_{x} \psi_1 \bar{S}(x,\tau)- \pa_{x} \bar{\psi}_1 S(x,\tau)\Big] \label{N0}
\er
The last equation, taking into account the source $S$ (\ref{source}) and the expressions of the dark solitons (\ref{para1})-(\ref{para2}), can be rewritten as
\br
\nonumber
{\cal N}(t) &=& 2 \int_{-\infty}^{t} d\tau \int_{-\infty}^{+\infty} \, dx \, \,  \frac{d \phi_1}{dx} \Big\{ \Big[ \phi_2 \sin{(\theta -\zeta)} + |\psi_0| \sin{(\zeta + \d)} \Big] f'[|\Psi_0|^2] - \\
& &  \phi_2 \sin{(\theta-\zeta)} f'[|\phi_2|^2] - |\psi_0| \sin{(\zeta+ \d)} f'[|\psi_0|^2]  \Big\},
\label{num11}\\
|\Psi_0|^2  &=& \phi_1^2 + \phi_2^2+ |\psi_0|^2 + 2 \phi_1 \phi_2 \cos{(\theta-\zeta)} - 2 \phi_1 |\psi_0| \cos{(\zeta+\d)} - 2 \phi_2 |\psi_0| \cos{(\theta-\d)}.\label{P0}
\er

In the next steps we will change the space-time $(x,\tau)$ integration to the field space $(\phi_1, \phi_2)$ integration in the limit $t \rightarrow +\infty$. So, consider the coordinate transformation 
\br
x' = x,\,\,\,\,z = (x- v \tau),\,\,\, \, dx d\tau = \frac{1}{ v} dx' dz  
\er
The above transformation and (\ref{ff1})-(\ref{ff2}) allow us to write $dz = \pm d\phi_2/G[\phi_2]$, where the minus (plus) sign corresponds to the left (right) interval of integration $\int_{|\psi_0|}^{q_0} d\phi_2...$ ($\int^{|\psi_0|}_{q_0} d\phi_2...$), respectively. Notice that $q_0$ is the minimum intensity of the gray soliton, and it is assumed that the gray soliton is approaching the localized region of the black soliton from the negative $x$ direction with a speed $v$. Therefore, in the limit $t \rightarrow +\infty$ the expressions for ${\cal N}(+\infty)$ and ${\cal D}(+\infty)$ in field space can be written as
\br
\nonumber
{\cal N} &\equiv & -\frac{4}{v} \int_{-|\psi_0|}^{|\psi_0|} d\phi_1 \int_{q_0}^{|\psi_0|}\frac{d \phi_2}{G[\phi_2] } \Big\{ \Big[ \phi_2 \sin{(\theta -\zeta)} + |\psi_0| \sin{(\zeta + \d)} \Big] f'[|\Psi_0|^2] - \\
& &  \phi_2 \sin{(\theta-\zeta)} f'[|\phi_2|^2] - |\psi_0| \sin{(\zeta+ \d)} f'[|\psi_0|^2]  \Big\}, \label{N1}\\ 
{\cal D} & \equiv  & 2 \int_{-|\psi_0|}^{|\psi_0|} d\phi_1  F[\phi_1]
\label{d11},
\er
where the field $\theta$ appearing above explicitly, as well as through the expression for $|\Psi_0|$ given in (\ref{P0}), must be substituted in terms of the `integration variable' $\phi_2$ through $\theta = \Theta[\phi_2]$ from (\ref{id1}). 

So, after the colliding shallow `gray' soliton $\psi_2$ is sufficiently far from the stationary `black' soliton $\psi_1$  the spatial shift of it  becomes 
\br
\label{dx111}
\D x = \frac{{\cal N}}{2{\cal M}},  \,\,\,\,\  \,\,\,\,{\cal M}\equiv \frac{\cal D}{2} \er
where ${\cal M} = \frac{1}{2}\int  dx  [ (\frac{d \phi_1}{dx})^2 + (F[\phi_1^2])^2 ]$ is the energy of the stationary black soliton.
 
For some models which allow analytic solitary wave solutions one observes that the maximum $|\psi_0|$ and minimum $q_0$ intensities of the incoming wave are associated, respectively, to some roots of the functional $G[\phi_2]$. This feature is present in the models with saturable nonlinearities \cite{kroli} and in the NLS and CQNLS models presented in the appendices \ref{dark} and \ref{cq1}, respectively.  Then, the integrand in (\ref{N1}) diverges in the both end points of the integration interval, requiring a certain regularization procedure in order to extract a finite value for the spatial shift expression. Moreover, the difference between the roots behave as $(|\psi_0|-q_0) \rightarrow 0$ for $v \rightarrow v_s$, i.e. in the sound speed limit. Making use of the above properties (discussed in the appendices for the NLS and CQNLS cases) and in order to see the appearance of the perturbative parameter $1/(v \g)$ in the expression for  ${\cal N}$  in (\ref{N1}) one makes a transformation to a convenient field space 
 \br
\label{trans1} d\phi_1 d\phi_2 = \frac{1}{(v \g)} \Phi'[y] d\phi_1 du,
\er with $\phi_2 \equiv \Phi[y]$, $y=u/\g$, $\g^{-1} = \sqrt{1-(\frac{v}{v_s})^2}$, such that (\ref{N1}) can be written as
\br
\label{ovfact}
{\cal N} &=& -\frac{4}{(v \g)} \int_{0}^{u_o} du \int_{-|\psi_0|}^{|\psi_0|} d\phi_1  {\cal Z}[\phi_1, u],\\
  &\equiv & -\frac{4}{(v \g)} \int_{0}^{u_o} du {\cal K}[u]\label{ovfact1}
\er
where
\br
{\cal Z}[\phi_1, u] &=&\frac{1}{G[\Phi[\frac{u}{\g}]] } \Big\{ \Big[ \Phi[\frac{u}{\g}] \sin{(\theta -\zeta)} + |\psi_0| \sin{(\zeta + \d)} \Big] f'[|\Psi_0|^2] -\nonumber \\
& &  \Phi[\frac{u}{\g}] \sin{(\theta-\zeta)} f'[(\Phi[\frac{u}{\g}])^2] - |\psi_0| \sin{(\zeta+ \d)} f'[|\psi_0|^2]  \Big\} \label{tt1}\\
|\Psi_0|^2  &=& \phi_1^2 + (\Phi[\frac{u}{\g}])^2+ |\psi_0|^2 + 2 \phi_1 \Phi[\frac{u}{\g}] \cos{(\theta-\zeta)} - 2 \phi_1 |\psi_0| \cos{(\zeta+\d)} - \nonumber \\ && 2 \Phi[\frac{u}{\g}] |\psi_0| \cos{(\theta-\d)}. \label{tt2} 
\er

The functional ${\cal K}[u]$ is obtained from (\ref{ovfact}) by integrating the functional ${\cal Z}[\phi_1, u]$ in $\phi_1$. The phase function $\theta$ which appears in the integration (\ref{ovfact1}) through (\ref{tt1})-(\ref{tt2}) must be written in terms of $u$ by making use of the relationship 
 $\theta = \Theta[\Phi[\frac{u}{\g}]]$ defined in (\ref{id1}), where the minus sign has already been taken into account in the relevant region of integration. Notice that the factor $1/(v \g)$ above is not directly obtained from a space time transformation as in the analog computations in relativistic scalar field theory models \cite{amin1, amin2}, in our case it also requires an additional field space transformation (\ref{trans1}). The parameter $u_o$ in general may depend on $v$ and the parameters of the theory. For example in the NLS case below we will find that $u_o$ is a vanishing parameter in the sound speed limit and in the CQNLS case the relevant $u_o$ is a constant equal to unity. So, making use of eq. (\ref{dx111}) and computing the zero'th order term of the integral in (\ref{ovfact1}) one can write the spatial shift as 
\br\label{ssf}
\D x = -\frac{2}{(v\g){\cal M}} \Big[\int_{0}^{u_o} du  {\cal K}[u] \Big]_{(0)}+ {\cal O}[(v \g)^{-2}],     
\er
where ${\cal M}$ is the energy of the black soliton and $[...]_{(0)}$ means that only the zero'th order term of the full integral is considered. Remarkably, in order to compute $\D x$ we need only the form of the functional $f[I]$ which enters into the relationships (\ref{ovfact})-(\ref{ssf}) explicitly and through the functional $G[\phi_2]$ as defined in (\ref{ff2}). Despite the fact that the collision is dissipative the spatial shift does not exhibit a time dependence implying that no velocity change is observed at leading  order. 

In this way the above result for the spatial shift furnishes the leading order term of the stationary soliton perturbation $h(x,t)$ through the expression (\ref{exp}). Note that in that expression this contribution accounts only for the zero mode related to space translation. 

In the construction above the {\sl continuous wave background} is assumed to be at rest, i.e. $k=0$ in (\ref{cwb}) and it is suitable for treating the shallow dark (gray)-black soliton collisions. A variant of the above method applied to high speed gray-gray soliton collisions requires a non-rest {\sl cwb}, as we will see below. We consider a small amplitude (shallow) dark soliton with velocity $v_2$ and a wide width $\frac{1}{p_2}$ colliding with a large amplitude gray soliton with velocity ($-v_1$) and a narrow width $\frac{1}{p_1}$, such that $v_2 > v_1$ ($v_1,v_2 >0$) and $p_2 < p_1$. Notice that in the integrable NLS case the analytic 2-dark soliton solution involves these parameters satisfying the relationship $2 k = v_2 - v_1 + \frac{p_2^2- p_1^2}{v_1+v_2}$, where $k$ is the wave number of the background (\ref{cw}). In order to compute the spatial shift experienced by the slow thin soliton after colliding with the fast broad soliton in the context of our formalism it is convenient to consider the co-moving coordinate of the low velocity thin soliton. In this frame, the large amplitude soliton is stationary, whereas the small amplitude soliton travels with a relative velocity, and the oscillatory {\sl cwb} is flowing with certain phase velocity. In this frame consider the {\sl cwb} solution of the deformed NLS model (\ref{nlsd}) as a non-rest oscillatory phase       
\br
\label{cwb1}
\hat{\psi}_0(x,t) = |\psi_0| e^{i \( k x + w t + \d\)},
\er
where $k$, $w = - \frac{1}{2} k^2- f'[|\psi_0|^2]$ and $w/k$ are the wave number, the dispersion relation and the phase velocity of the background, respectively. A stationary soliton can be transformed into a nonstationary form by using the Galilean symmetry. In fact, assuming that $\psi(x,t)$ is a solution of the deformed NLS equation (\ref{nlsd}) and $c$ a constant, then $\psi'(x,t) \equiv e^{i(c \,x -\frac{c^2}{2} t)} \psi(x - ct, t)$ will also be a solution. So, consider a solution before, during and after the collision of gray-gray solitons in the co-moving frame of the slow soliton as  
\br \label{total11} \Psi(x,t) &\equiv & \psi_{1}(x,t) + \psi_{2}(x,t) + \hat{h}(x,t) - \hat{\psi}_0(x,t)\\
 &\equiv & \phi_1(x) e^{i \theta_1(x)} e^{i \( k x + w t + \zeta_1\)} + \phi_2(x- v t) e^{i \theta_2(x - v t )} e^{i \( k x + w t + \zeta_2\)} + \hat{h}(x,t) -\nonumber \\
  && |\psi_0| e^{i \( k x + w t + \d\)} \label{total22}.\er
Here $\psi_1(x,t)$ represents the stationary gray soliton and $\psi_2(x,t)$ the moving gray soliton such that $v$ is the relative velocity between the soliton and the background. The constant phase parameters $\zeta_1, \zeta_2, \d$ have been included. The function $\hat{h}(x,t)$ describes the perturbation effects around the core of the stationary gray  soliton generated by the collision in the co-moving frame. As above it has been included the {\sl cwb} (\ref{cwb1}) in order to subtract the {\sl nvbc} reference term  $\hat{\psi}_0(x,t)$ on top of the superposition. Following similar steps as in section \ref{solwaves} one can obtain the relationships between the functions $\phi_1(x)$ and $\theta_1(x)$, as well as between  $\phi_2(z)$ and $\theta_2(z),\,(z= x- v t)$, valid in the co-moving frame. The expansion in (\ref{total11}) can be substituted into the deformed NLS equation (\ref{nlsd}) in order to get a linearized equation for $\hat{h}(x,t)$ as in eqs. (\ref{eqh})-(\ref{d44}). The perturbative computations must be performed in the co-moving frame and they can follow the relevant steps as above; the stationary gray soliton can be Taylor expanded and related to the function $\hat{h}(x,t)$ as in eqs. (\ref{taylor}) and (\ref{exp}), respectively. Analogous expressions to the ones in eqs. (\ref{delta})-(\ref{P0}) can directly be reproduced. The relevant space-time integration as in (\ref{num11}) would be transformed to the field space integration: $(x, \tau) \rightarrow (\phi_1, \phi_2)$ in the limit $t\rightarrow +\infty$. A convenient transformation in the field space such that $d\phi_1 d\phi_2 \sim  \frac{1}{(v \g)} d\phi'_1 d\phi'_2$ can be implemented as in (\ref{trans1}). However, in the above construction some integrals would require a renormalization procedure in order to remove the divergences associated to the contribution of the oscillatory $cwb$ \cite{kivshar}. For example, the integral (\ref{dxt2}) becomes in this case ${\cal D} = 2 \int_{-\infty}^{+\infty} dx \{ [\frac{\pa}{\pa x} \phi_1(x)]^2+ \frac{k^2 |\psi_0|^4}{\phi_1^2(x)}\}$, which is divergent for $k\neq 0$ since $\phi_1 \rightarrow |\psi_0|$ as $|x|\rightarrow +\infty$. The renormalized expression ${\cal D}_{r} \equiv 2 \int_{-\infty}^{+\infty} dx \{ [\frac{\pa}{\pa x} \phi_1(x)]^2+ k^2 |\psi_0|^4[ \frac{1}{\phi_1^2(x)}-\frac{1}{|\psi_0|^2}]\}$ must be considered. An analogous expression to (\ref{ssf}) will provide the leading order $(v \g )^{-1}$ contribution to the spatial shift. 

Next, we present some examples of application of the above formalism to the gray-black soliton collisions, paying special attention to the relevant field space transformation in each model in order to extract the expansion parameter $1/(v\g)$.       
   
\section{Examples}

\label{sec4}

\subsection{NLS model}
Let us consider the integrable defocusing NLS model in order to test the formalism developed above. One has 
the NLS model defined by   
\br
\label{nls2}
i \frac{\pa}{\pa t} \psi(x,t) + \frac{1}{2} \frac{\pa^2}{\pa x^2} \psi(x,t) - \b |\psi(x,t)|^2 \psi(x,t) =  0,\er
where we have considered  $f'[I] = \b I$ in (\ref{nlsd}). Note  that in this model a closed analytic expression is available for the spatial shift of colliding gray-black solitons (\ref{shiftnls1})- (\ref{shiftnls2}). Next let us compute the spatial shift applying our formalism. So, take 
\br
\label{nls0}
f[I] = \frac{\beta}{2} I^2.
\er

Then, from the eqs. (\ref{ff1})-(\ref{id1}) one can get
\br
F[\phi_1] &=&  \sqrt{\b} (|\psi_0|^2-\phi_1^2),\,\,\,G[\phi_2] = \frac{|\psi_0|^2-\phi_2^2}{\phi_2} \sqrt{\b \phi^2_2 - v^2},\,\,\, H[\phi_2] = v (1-\frac{|\psi_0|^2}{\phi_2^2}) ,\\ \theta &=& \pm \arcsin{\( \frac{v}{\sqrt{\beta}} \frac{1}{\phi_2}\)}-\theta_{1}, \label{theta2}
\er
where $\theta_1$ is a constant of integration. The function $G[\phi_2]$ possesses two roots, i.e. $\{|\psi_0|,\,\,v/\sqrt{\b}\}$. The first root is the maximum intensity, whereas the second one is the minimum intensity (dip) of the dark soliton, respectively.
In the field space integration of ${\cal N}$ (\ref{N1}) the field $\phi_1$ appears only in the expression of $f'[|\Psi_0|^2] = \b |\Psi_0|^2$, where $|\Psi_0|^2$ is given in (\ref{P0}); so after integrating in $\phi_1$ one has
\br
\nonumber
{\cal N} = -\frac{4 |\psi_0| \b}{v} \int^{|\psi_0|}_{\frac{v}{\sqrt{\beta}}} \frac{d \phi_2}{G[\phi_2]} \Big\{ \Big[ \phi_2 \sin{(\theta-\zeta)} + |\psi_0|  \sin{(\zeta+\d )}  \Big] \Big[ \frac{4 |\psi_0|^2}{3}+ \phi_2^2  - \\
2  |\psi_0|  \phi_ 2 \cos{(\theta - \d)} \Big] - \phi_2^3 \sin{(\theta-\zeta)} -  |\psi_0|^3 \sin{(\zeta+ \d)}\Big\}, \label{NN1}
\er
where the relationship (\ref{theta2}) must be used in order to write the integrand only in terms of the `integration variable' $\phi_2$. The above expression can be written in terms of the perturbation parameter $(\frac{1}{v \g})$ as in (\ref{ovfact}) in order to organize the perturbative series. In fact, the expression (\ref{NN1}) can be written as 
\br
\nonumber
{\cal N} &=& - \frac{4 v_s}{(v \g)} \int_{0}^{u_o} \frac{d u}{u_o^2- u^2} \Big\{ \Big[ \phi_2[u] \sin{(\theta-\zeta)} + |\psi_0|  \sin{(\zeta+\d )}  \Big] \Big[ \frac{4 |\psi_0|^2}{3}+ \phi_2^2 [u] - \\
&& 2  |\psi_0|  \phi_ 2[u] \cos{(\theta - \d)} \Big] - \phi_2^3[u] \sin{(\theta-\zeta)} -  |\psi_0|^3 \sin{(\zeta+ \d)}\Big\}, \label{NN11}\\
\phi_ 2[u] & \equiv & \sqrt{\frac{v^2}{\b} + \g^2 u^2},\,\,\,\,u_o \equiv \frac{v_s^2 -v^2}{|\psi_0| \b}.
\er 
Notice that the upper limit of the integration interval $u_o$ approaches zero in the sound speed limit $v \rightarrow v_s$. So, due to the overall factor $(\frac{1}{v \g})$ it can be sufficient to evaluate the zeroth order term of the integral. However, it is instructive to perform the full integration in order to see the nature of the divergences. In fact, the integrand in (\ref{NN1}) presents certain divergences at the both ends of the interval $[\frac{v}{\sqrt{\b}}, |\psi_0|]$. They are associated to the roots of the functional $G[\phi_2]$ in the denominator, so let us examine the relevant divergent terms by performing the full integration. The associated indefinite integral in (\ref{NN1}), after integration, becomes 
\br\nonumber
&&\frac{|\psi_0|}{6\b} \left\{6 v \cos{(2 \theta_1+\zeta +\delta)}+ \sqrt{\beta } |\psi_0|  \Big[4 \sin{(\zeta+\theta_1)} +3 \sin{(\theta_1+\zeta +2 \delta)} - 3 \sin{(\theta_1-\zeta)}\Big]\right\} \times\\ \nonumber
&& \log{(|\psi_0|^2-\phi_2^2)}+\\
&&\frac{|\psi_0|}{3\b \sqrt{v_s^2-v^2}} \Big[ 4 v \sqrt{\b} |\psi_0| \cos{(\zeta+\theta_1)} -  3 v \sqrt{\b} |\psi_0| \cos{(\theta_1-\zeta)}+3 \sqrt{\b} v |\psi| \cos{(\zeta+\theta_1+2\d)} + \nonumber \\
&& 3 \b |\psi_0|^2 \sin{(\zeta-\d)}+4 \b |\psi_0|^2 \sin{(\zeta+\d)} -
 6 v^2  \sin{(2\theta_1+\zeta+  \d)}+3 \b |\psi_0|^2 \sin{(2\theta_1 + \zeta+ \d)} \Big] \times \nonumber \\
&&\mbox{arctanh}\(\frac{\sqrt{\b \phi_2^2 - v^2}}{\sqrt{v_s^2-v^2}}\)-\nonumber\\
&&\frac{|\psi_0|}{\b} \sqrt{\b \phi_2^2-v^2} \Big[ \sin{(\zeta-\d)} + 2\cos{\theta_1} \sin{(\theta_1+\zeta+\d)}\Big]. \label{finite1}
\er

The above expression, except the finite term of the last line, presents certain divergences when evaluated at the end points of the integration interval, i.e. the functions $\log{(|\psi_0|^2-\phi_2^2)}$ in the second line and $\mbox{arctanh}\(\frac{\sqrt{\b \phi_2^2 - v^2}}{\sqrt{v_s^2-v^2}}\)$ in the fifth line diverge at the upper and lower limits of the integration interval, respectively. In order to left with the finite value of the last line in the  expression (\ref{finite1}) we can choose the  arbitrary parameters $\zeta,\, \d,\,\theta_1$ in such a way that the coefficients of these divergent functions vanish identically. So, let us assume the following relationships between the parameters
\br
\label{rel11}
6 v \cos{(2 \theta_1+\zeta +\delta)}+ \sqrt{\beta } |\psi_0|  \Big[4 \sin{(\zeta+\theta_1)} +3 \sin{(\theta_1+\zeta +2 \delta)} - 3 \sin{(\theta_1-\zeta)}\Big]=0,\,\,\,\,\\
  4 v \sqrt{\b} |\psi_0| \cos{(\zeta+\theta_1)} -  3 v \sqrt{\b} |\psi_0| \cos{(\theta_1-\zeta)}+3 \sqrt{\b} v |\psi_0| \cos{(\zeta+\theta_1+2\d)} + \nonumber \\
 3 \b |\psi_0|^2 \sin{(\zeta-\d)}+4 \b |\psi_0|^2 \sin{(\zeta+\d)} -
 6 v^2  \sin{(2\theta_1+\zeta+  \d)}+3 \b |\psi_0|^2 \sin{(2\theta_1 + \zeta+ \d)}  =0.\,\,\,\,\label{rel22}
\er

From the system (\ref{rel11})-(\ref{rel22}) one gets the relationships
\br\label{the1}
\theta_1 &=& -\zeta -\d - \frac{1}{2} \arccos{(2/3)},\\
\label{v11} 
\zeta &=& \arccos{\Big[\frac{\frac{v}{v_s} - \sqrt{5} \sqrt{1-(\frac{v}{v_s})^2}}{\sqrt{6}}\Big]}.
\er

Therefore, using the relations (\ref{rel11})-(\ref{rel22}) in the expression (\ref{finite1})  one gets  a  finite value for ${\cal N}$ in (\ref{NN1}) 
\br
{\cal N} = - \frac{4 |\psi_0|^2}{v} \sqrt{v_s^2-v^2} \, \Big[ \sin{(\zeta-\d)} + 2\cos{\theta_1} \sin{(\theta_1+\a+\d)}\Big],
\er
where the parameters $\zeta$ and $\theta_1$ can be written in terms of $v, v_s , \d$ through the relationships (\ref{the1})-(\ref{v11}). 

Then, taking into account ${\cal D} = 2 {\cal M}_{NLS} = \frac{8}{3} \sqrt{\beta} |\psi_0|^{3}$ from (\ref{d11}) and the above expression for ${\cal N}$ written in terms of the parameters $\{\g, \d, v\}$, one gets the spatial shift from the equation (\ref{dx111})
\br
\label{low0}
\D x_{pert.} &=& -\frac{1}{\g v}  \Big[\frac{A(\g) \cos{\d} + B(\g) \sin{\d}}{\sqrt{6} \g}\Big],\\
A(\g)&\equiv& -\sqrt{5} \sqrt{\gamma ^2-1}+4 \sqrt{-4+5 \gamma ^2+2 \sqrt{5} \sqrt{\gamma ^2-1}}+5,\\B(\g) &\equiv& -2 \sqrt{\gamma ^2-1} +\sqrt{5} \sqrt{-4 + 5 \gamma^2 + 2 \sqrt{5} \sqrt{\gamma^2-1}}+2 \sqrt{5}\\ 
  \g &\equiv& \(1-\frac{v^2}{v_s^2}\)^{-\frac{1}{2}}, \,\,\,\, v_s \equiv \sqrt{\b} |\psi_0| ,  , \label{param11} 
\er 
where $v_s$ is the sound speed. Notice that the spatial shift $\D x$ will depend on three independent parameters, $\{v, v_s, \d\}$. However, the parameter $\d$ can be fixed by considering the convergence of our series expansion to the value provided by the exact NLS phase shift in the limit $v \rightarrow v_s$.  
 
On the other hand, the exact analytic expression for the spatial shift experienced by a black soliton after colliding with a gray soliton in the NLS model becomes (see (\ref{shiftnls1})-(\ref{shiftnls2})) 
\br
\label{analy}
\D x_{analytic} & = & \frac{1}{2 \,v_s} \log{\frac{\g - 1}{\g + 1}}\\
 &=& - \frac{1}{\g v} + {\cal O}[(\g v)^{-2}]\label{analy1}
\er
Whereas, the leading order term of the expression (\ref{low0}) in the limit
 $v \rightarrow v_s\, (\g >>1)$ becomes 
\br \D x_{pert.} =  - \frac{(3/2)}{\g v}  \(\frac{\sqrt{5} \cos{\d} + \sin{\d}}{\sqrt{6}}\)+ {\cal O}[(\g v)^{-2}]\label{lim1}. \er

Comparing the both expressions (\ref{analy1}) and (\ref{lim1}) one gets $\d  = \arccos{\sqrt{\frac{5}{54}}}$. Then
 \br \D x_{NLS} =  - \frac{1}{\g v} + {\cal O}[(\g v)^{-2}]\label{lim2}. \er

Therefore, our expression (\ref{low0}) reproduces correctly the first order approximation in the parameter expansion $1/(\g v)$ of the spatial shift experienced by a black soliton after colliding with an shallow gray soliton in the NLS model. Let us evaluate the analytical results for some values of the parameters. The Fig. 1 shows  a plot of the gray-black soliton collision in which the exact spatial shift using (\ref{analy}) becomes $\D x_{analy} = - 0.12437$ which must be compared to the perturbation theory result $\D x_{NLS} \approx -0.1364$, then this result provides an approximation within an error of $9 \% $. This is acceptable since the velocity of the gray soliton being $v=4.64 $  ($v_s=6,\,v < v_s $) one has  $ (v \g) = 7.33$, which is not expected to be a parameter characterizing an shallow soliton.  The Fig. 4 shows the spatial shift $\D x$ vs $(v\g)$ for various values of the sound speed $v_s=1,\,6,\,12$. The continuous lines correspond to the exact analytical result (\ref{analy}) for $v_s=1$(green),$v_s= 6$(blue) and $v_s=12$ (red), respectively. The theoretical prediction (\ref{lim2}) is plotted as a dashed line. The dashed and green lines exhibit an excellent agreement for $(v\g) > 3$.  The inner plot shows that the expression in (\ref{lim2}) exhibits  a good convergence even for values $(\g v) \approx 3$ provided that the sound speed satisfies $v_s \leq 1$. Therefore, we can conclude that the leading order result (\ref{lim2}) is in excellent agreement with the exact result of $\D x$ provided that $v_s \leq 1$. Moreover, using the expression (\ref{param11}) of $v_s$ in terms of the parameters of the theory one can argue that   a good convergence of the leading order result provided by (\ref{lim2}) is restricted to a region in parameter space such that $|\psi_0| \leq 1/\sqrt{\b}$.
 
The behavior of the dark solitons in the sound speed limit (i.e. when the solitons become infinitely shallow, such that they reduce effectively to a pure plane wave) can be studied by considering it as being equivalent to the Bogoliubov phonon, so the Bogoliubov-de Gennes equation (BdG) (\ref{bdg}) associated in this case to the integrable NLS model can be considered as a scattering problem with the static black soliton potential. It is known that the dark soliton does not scatter phonons, i.e. the scattering of the phonon field with dark soliton potentials is reflectionless \cite{chen, dziarmaga}. This behavior has also been observed in BdG type equations with soliton potentials related to some integrable systems (see \cite{koller} and references therein). This feature implies that the spatial shift of the black soliton after colliding with an extremely shallow soliton must tend to zero in the sound speed limit. This behavior is consistent with our result above in this limit.  

The low velocity ($v < \frac{v_s}{2}$) NLS soliton collisions have recently been studied by considering the solitons as hard-sphere-like particles which interact through an effective repulsive potential \cite{theocharis}. The validity of the method has been checked by comparing to the exact analytical results for the collision-induced phase-shifts of the solitons, showing an excellent agreement in this regime. This analysis of soliton interactions has also been used to study a particular deformation of NLS model (\ref{nlsd}) ( $f'[I] = \sqrt{1+ 4 I}$ ) which considers an additional external harmonic potential \cite{theocharis}. The relevant theoretical results are in good agreement with the corresponding  numerical simulations. In fact, at low velocities the solitons do not overlap completely during collision, instead they reflect from each other from the point of their closest proximity, so that they can be characterized by two individual density minima even at the collision region. However, high speed solitons ($v > \frac{v_s}{2}$) are transmitted through each other, so they overlap completely in the interaction region  and approach a single density minimum characterizing the location of the both solitons. This property has been used in our analysis above related to the regime in which the incoming shallow dark soliton possesses high speed and transmits through the stationary black soliton.           

\subsection{Away from NLS: cubic-quintic non-linear Schr\"{o}dinger (CQNLS)}

Consider the non-integrable cubic-quintic NLS model defined by
\br   
\label{cqnls2}
i \frac{\pa}{\pa t} \psi(x,t) + \frac{1}{2} \frac{\pa^2}{\pa x^2} \psi(x,t) -\( \b |\psi(x,t)|^2 - \frac{\a}{2} |\psi(x,t)|^4\) \psi(x,t) =0,\,\,\,\, \b > 0, \a > 0,
\er
where we have considered $ f'[I] = \b I - \frac{\a}{2} I^2$ in (\ref{nlsd}).
In the Appendix \ref{cq1} we present the dark-soliton type solutions of this model, as well as the relevant expressions to be used in the equations  (\ref{N1})-(\ref{ssf}) in order to compute the phase shift. So, let us compute the spatial shift experienced by a static black soliton (\ref{bl1}) after colliding with a shallow gray soliton (\ref{sl1}). In the field space integration of ${\cal N}$ (\ref{N1}) the field $\phi_1$ appears only in the expression of $f'[|\Psi_0|^2] = \b |\Psi_0|^2-\frac{\a}{2} |\Psi_0|^4 $, where $|\Psi_0|^2$ is given in (\ref{P0}); so using the  relationship (\ref{G2}) and substituting the field $\theta$ in terms of $\phi_2$ from (\ref{relation1}), the expression (\ref{N1}) can be written as
\br
{\cal N} =- \frac{2}{v} \sqrt{\frac{3}{\a}} \int^{|\psi_0|}_{\sqrt{\xi_1}}  \frac{{\cal K}[\phi_2,\zeta,\theta_1, \delta]\,  \phi_2 \, d\phi_2}{(|\psi_0|^2-\phi_2^2) \sqrt{(\phi_2^2-\xi_1)(\xi_2-\phi_2^2)}}, \label{NN11c}
\er
where the integration in $\phi_1$ in the interval $[-|\psi_0|,|\psi_0|]$ has already been performed and the integration limits of $\phi_2$  is defined as the minimum and maximum intensities of the incoming dark soliton, i.e.  $[\sqrt{\xi_1},|\psi_0|]$ (see Appendix \ref{cq1}). The functional ${\cal K}$ becomes 
\br
\nonumber
{\cal K}[\phi_2,\zeta,\theta_1, \delta ] &\equiv & \frac{|\psi_0| \phi_2^2(\a \phi_2^2-2 \b)}{\sqrt{\xi_2-\xi_1}} C + |\psi_0|^4(\a |\psi_0|^2-2 \b) \sin{(\zeta+\d)}-\frac{|\psi_0|}{15 \sqrt{\xi_2-\xi_1}}
\Big\{ \a (38 |\psi_0|^4 + \\ \nonumber && 80 |\psi_0|^2 \phi_2^2 + 15 \phi_2^4)-10 \b (4 |\psi_0|^2 + 3 \phi_2^2) + 10 |\psi_0| \Big[ \a |\psi_0|^3 \cos{[2 (\zeta + \d])} + \\
\nonumber && \frac{ \a |\psi_0| B_1 + 3 \a |\psi_0| B_2}{\xi_1-\xi_2} +  \frac{(3 \b - 4 \a |\psi_0|^2 - 3 \a \phi_2^2)2 A_2 - 4 \a A_1|\psi_0|^2 \cos{(\zeta + \d)}}{\sqrt{\xi_2-\xi_1}} \Big] \Big[ C + \\ 
&& \sqrt{\xi_2 - \xi_1} |\psi_0| 
\sin{(\zeta + \d)} \Big]\Big\},\label{functional}\\\nonumber
A_1 &\equiv &  \sqrt{\xi_2-\phi_2^2} \sqrt{\xi_1} \cos{(\theta_1 - \zeta)} + 
 \sqrt{\phi_2^2-\xi_1} \sqrt{\xi_2} \sin{(\theta_1 - \zeta)}\\\nonumber
A_2 &\equiv &  \sqrt{\xi_2-\phi_2^2} \sqrt{\xi_1} \cos{(\theta_1 - \d)} + 
 \sqrt{\phi_2^2-\xi_1} \sqrt{\xi_2} \sin{(\theta_1 - \d)}\\
 \nonumber
 B_1  &\equiv &  [\xi_2 \phi_2^2 + \xi_1 (\phi_2^2-2\xi_2) ] \cos{[2(\theta_1 - \zeta)]} -2 
 \sqrt{(\phi_2^2-\xi_1)(\xi_2-\phi_2^2)} \sqrt{\xi_2\xi_1}  \sin{[2(\theta_1 - \zeta)]}
\\
 \nonumber
 B_2  &\equiv &  [\xi_2 \phi_2^2 + \xi_1 (\phi_2^2-2\xi_2) ] \cos{[2(\theta_1 - \d)]} -2 
 \sqrt{(\phi_2^2-\xi_1)(\xi_2-\phi_2^2)} \sqrt{\xi_2\xi_1}  \sin{[2(\theta_1 - \d)]}\\
 \nonumber
C &\equiv &  -\sqrt{\phi_2^2-\xi_1} \sqrt{\xi_2} \cos{(\theta_1 - \zeta)} + 
 \sqrt{\xi_2-\phi_2^2} \sqrt{\xi_1} \sin{(\theta_1 - \zeta)}.
\er 
Notice that in order to find the relationship (\ref{relation1}) between the fields $\phi_2$ and $\theta$ it is not necessary to have the explicit analytical solution of the gray soliton, since that relationship can be obtained from (\ref{ff2})-(\ref{id1}). 
It is clear that the integral in (\ref{NN11c}) diverges at the end  points of the integration interval which are associated to the roots of the functional $G[\phi_2]$ (\ref{G2}). In order to extract a finite value for this integral, as in the computation of the usual NLS case above, one could examine the relevant divergent terms of the indefinite integral and choose convenient values for the set of parameters $\{\zeta, \d, \theta_1\}$ such that the definite integral becomes finite. However, in the present case this procedure turns out to be very cumbersome since it involves various types of Elliptic integrals and the relationship between the parameters would be complicated, therefore we will perform the computation considering a power series expansion in the perturbation parameter $1/(v\g)$. 

So, we make a relevant transformation in order to write (\ref{NN11c}) in the form (\ref{ovfact})-(\ref{ovfact1}). Taking the following transformation
\br
u = \frac{\g}{v_s} \sqrt{\a/3} \sqrt{(\phi_2^2-\xi_1)(\xi_2-\phi_2^2)},
\er
the expression (\ref{NN11c}) becomes
\br
{\cal N} &=&-\frac{3 v_s}{\a} \, \(\frac{1}{ v \g}\)\, \int_{0}^{1}  \frac{{\cal K}[\phi_2[u],\zeta,\theta_1, \delta]\,\, dy}{[|\psi_0|^2-(\phi_2[u])^2] [\rho^{+}-(\phi_2[u])^2]},\label{NN22}\\
(\phi_2[u])^2  & \equiv & \rho^{+} + \rho^{-} \Big[1+ \( \frac{3 v_s^2}{\a (\rho^{+})^2}\) \, \frac{1-u^2}{\g^2} \Big]^{1/2},
\er
where $\rho^{\pm} \equiv \frac{1}{2} |\psi_0|^2 \pm \frac{3}{2 \a} \frac{v_s^2}{|\psi_0|^2}$. Remarkably, the overall factor $1/(v \g)$, which is the expansion parameter in the problem, appears in the  expression (\ref{NN22}). Then it will be sufficient to compute the zeroth order term of  the integral in the power series expansion in the parameter $1/(v\g)$. However, this process must be performed  carefully due to the divergence of the integrand at $u=1 (\phi_2 = |\psi_0|)$ which appears explicitly in (\ref{NN22}) and the indeterminate values which may appear when some terms of the functional ${\cal K}[\phi_2,\zeta,\theta_1, \delta]$ vanishes at this point. Notice that this divergent point is also achieved in the sound speed limit $\frac{1}{v \g} \rightarrow 0,\, \xi_1 \rightarrow |\psi_0|,\, \xi_2 \rightarrow \frac{3 v_s^2}{\a |\psi_0|^2}, \phi_2 \rightarrow |\psi_0|$ (see (\ref{limit22})). In fact, the functional ${\cal K}[\phi_2,\zeta,\theta_1, \delta]$ in (\ref{functional}) contains some terms proportional to $\sqrt{\phi_2^2-\xi_1}$ which vanishes in the sound speed limit.

Since we are interested in the sound speed limit  of the integral in (\ref{NN22}) let us write all the parameters as the series expansion $ x = x_0 + \frac{x_1}{\g} + \frac{x_2}{\g^2}+ ... \( x \equiv \{\d,\theta_1,\zeta,\xi_1,\xi_2\}\) $ and consider the series  expansion of the integrand as  $ {\cal H}[\phi_2[u],\zeta,\theta_1, \delta, \g ] \equiv  \frac{{\cal K}[\phi_2[u],\zeta,\theta_1, \delta]}{[|\psi_0|^2-(\phi_2[u])^2] [\rho^{+}-(\phi_2[u])^2]} $ in powers of   $1/(\g)$  
\br
{\cal H}[\phi_2[u],\zeta,\theta_1, \delta, \g ] = {\cal H}[\phi_2[u=1],\zeta_0,\theta_{10}, \delta_0, \g \rightarrow \infty ] + {\cal H}'[\phi_2[u=1],\zeta_0,\theta_{10}, \delta_0, \g\rightarrow \infty] \, (\frac{1}{\g}) + ...\label{series1}
\er 
where ${\cal H}'[\phi_2[u=1],\zeta_0,\theta_{10}, \delta_0, \g\rightarrow \infty] =\frac{\pa}{\pa \G }{\cal H}[\phi_2[u=1],\zeta_0,\theta_{10}, \delta_0, \g\rightarrow \infty],\,\,\G \equiv  1/\g$. 

We are interested in the zeroth order term of the series expansion above. For simplicity in the computations we will choose some set of parameters $\{\d, \theta_1,\zeta\}$ in the sound speed limit ($\g \rightarrow \infty$). In fact, a huge simplification  is achieved for the choice 
\br
\label{phases1}
\zeta= -\pi/2+{\cal O}[\g^{-2}],\,\,\,\,\theta_1 =0+{\cal O}[\g^{-2}],\,\,\,\,  \d=0+{\cal O}[\g^{-2}].
\er 
For the value $\theta_{1}\approx 0$ the phase $\theta(z)$ of the gray soliton will remain as an odd function in the moving coordinate system, up to the lowest order (see the discussion below (\ref{th1})). Moreover, the set of  values  (\ref{phases1}) renders finite the lowest order term of the integration in (\ref{NN22}). Considering the zeroth order contribution from (\ref{series1}) and performing  the integration in (\ref{NN22}), one has
\br
{\cal N} &=& -\( \frac{1}{v \g} \) \times  \label{NN3}\\
&& \frac{8 \a^{3/2} |\psi_0|^3 \sqrt{\b - \a |\psi_0|^2} \( 45 \b^2 - 
   110 \a \b |\psi_0|^2 + 63 \a^2 |\psi_0|^4\)}{5 \sqrt{3} (3 \b - 4 \a |\psi_0|^2 )^2 [ \sqrt{\a} \b |\psi_0| \sqrt{3} \sqrt{\b - \a |\psi_0|^2} -(\b - 2\a |\psi_0|^2)(3\b - 2\a |\psi_0|^2) \mbox{tan}^{-1}(\frac{\sqrt{a} |\psi_0|}{\sqrt{3} \sqrt{\b-\a |\psi_0|^2}})]} 
    \nonumber
\er

From (\ref{MCQ}) one gets ${\cal D} = 2 {\cal M}_{CQ}$ and therefore the spatial shift, using the relationship (\ref{dx111}) becomes
\br
\D x_{CQNLS} &=&\(\frac{1}{v \g}\) \frac{8 \l^{3/2} \sqrt{1-\l} (45-110 \l +63 \l^2 )}{5\sqrt{3} (3-4 \l )^2 [(3-8\l+ 4 \l^2) \arctan{(\frac{\sqrt{\l}}{\sqrt{3} \sqrt{1-\l}})}-\sqrt{3} \sqrt{\l (1-\l)} ]},\,\,\,\,\,  \l \equiv \frac{\a |\psi_0|^2}{\b} \nonumber\\
&\equiv& -\frac{c(\l)}{v \g},\,\,\,\,\,\,\,\,\,\,0\leq \l <0.75.\label{shiftcqnls}
\er

This is the leading order in $\frac{1}{v \g}$ expression for the spatial shift experienced by a black soliton after collision with the relevant gray soliton in the CQNLS model. Some comments are in order here. First, notice the appearance of the parameter $\l$, which is defined as a relation between the quintic coupling $\a$ over the cubic coupling $\b$ constants multiplied by the squared amplitude $|\psi_0|^2$. In this way up to the first order in $1/(v \g)$ the spatial shift $\D x$ does not depend on the individual parameters of the theory, rather on the effective parameter $\l$ which is a function of them. The Fig. 5 shows  $\D x_{CQNLS}$ for various values of the effective parameter $\l$. Second, the above spatial shift $\D x$ reverses its sign from negative to positive values at the point $\l =0.654221$ as it is shown in Fig. 8. This implies that there are some sets of values of the parameters where $\D x =0$. Third, in the limit $\l \rightarrow 0$ one has  $\D x = - \frac{(3/2)}{v \g}$ ($c(0)= 3/2$), i.e. this is the spatial shift in the sound speed limit when the parameters satisfy $\a |\psi_0|^2 << \b$. There is another value of $\l$ for  which one has the same value of $c(0)$, i.e. $\l = 0.526644$ (the spatial shift for these values is plotted as a dashed line in Fig. 5). Moreover, the expression $c(\l)$ above has a singular point at $\l=0.75$, then one must have ($0\leq \l <0.75$).  Therefore, our leading order result is meaningful in the regions of parameter space such that $\a < \frac{3 \b}{4 |\psi_0|^2}$. Fourth, the study of the relationship between the rate of convergence of the first order approximation in (\ref{shiftcqnls}) and a set of parameters $\{\a,\b,|\psi_0|\}$ is interesting in its own right. The convergence check at a fixed $(v\g)$ can be performed by comparing (\ref{shiftcqnls}) and its $\l$ dependence to the numerically calculated spatial shift. For a deformed NLS model this rate of convergence problem can be related to the study of the functional form of $f[I]$ in eq. (\ref{nlsd}) and deserves a further exploration.

Next, we present a brief discussion about the properties of the shallow CQNLS dark solitons. In ref. \cite{crosta} it has been shown that the dark solitons become infinitely shallow at their bound for existence and that they are stable against weak perturbations. The problem of an infinitely shallow dark soliton and the outcome of its collision with the relevant black soliton can be discussed, as in the integrable NLS case above, in the context of  a scattering problem of the Bogoliubov phonon and the static black soliton potential associated to the Bogoliubov-de Gennes equation (\ref{bdg}) relevant to the CQNLS model. Recall that in the case of the BdG equation related to the integrable NLS model the phonon excitations are completely reflectionless against a dark soliton for any excitation energy. However, when a modified   nonlinear term is present, the phonon acquires a nonzero reflection coefficient in general.  This problem has been considered in \cite{takahashi} for the NLS model with general nonlinearity (\ref{nlsd})  and it has been shown that the property of perfect transmission of zero-energy phonon holds for a wide range of values of the velocity of the dark soliton which is used as the BdG equation potential. However, when a soliton velocity approaches a critical velocity this property disappears. In particular, a static black soliton does not scatter the zero-energy phonon. This feature implies that the spatial shift of the CQNLS black soliton after colliding with an infinitely shallow soliton must tend to zero. This behavior is consistent with our result above for the CQNLS model.

\section{Simulations}
\label{sec5}

The deformed NLS model (\ref{nlsd}) corresponds in general to a non-integrable model, so in the cases in which a dark soliton type solution exist in analytical form we do not expect to find an analytical expression for the collision of two solitons. That is the case for example in the cubic-quintic  and certain cases of the NLS with saturable nonlinearities. However, the analytical expression of one solitary wave can still  be used to provide an initial condition for the interaction of two solitary waves. So, in those cases we can take as the initial condition two one dark solitons some distance apart. 

After the fast-moving gray soliton has moved away from the region occupied by the stationary black soliton, one is left with the stationary black soliton profile plus perturbations generated by the collision process. The spectrum of perturbations includes the spatial shift $\D x$ of the position of the black soliton which we will compute numerically. The spatial shift can then be obtained by projecting the numerically calculated  $\widetilde{h}(x, t)$ onto $\pa_x \psi_1(x)$ (and their relevant complex conjugates), so from (\ref{exp}) one can write  
\br 
\label{numdx}
 \widetilde{\D x}(t) = \frac{1}{2}\frac{\int \Big[\widetilde{h}(x, t) \pa_x  \psi^{\star}_1(x) + \widetilde{h}^{\star}(x, t) \pa_x \psi_1(x)\Big]dx}{\int \pa_x \psi_1(x) \pa_x \psi^{\star}_1(x)dx},
\er
where the stars stand for complex conjugation and   $\widetilde{h}(x,t)$ can be obtained from (\ref{total}) as
\br \widetilde{h}(x,t) &\equiv & \widetilde{\Psi}(x,t)-\psi_1(x)-\psi_2(x,t)+\psi_0(x,t)\\
&=&\widetilde{\Psi}(x,t)-\phi_1(x) e^{i w t} e^{i \zeta} + |\psi_0| e^{i wt } (e^{i \d}-e^{i \theta^{-}}). \er

The function $\widetilde{\Psi}(x,t)$ is provided by the numerical evolution of the full equation. We compute $\widetilde{\D x}(t)$ in (\ref{numdx}) after the fast moving gray soliton has already been collided and it is sufficiently far from the stationary black soliton, then the asymptotic value of  the gray soliton has been taken as $|\psi_0| e^{i \theta^{-}} e^{i wt}$. It has been  considered that the dark soliton is approaching from the negative direction of $x$, then after the collision one has the asymptotic value of the phase  $\theta(-\infty)=\theta^{-}$. We consider a finite integration range in (\ref{numdx}) in such a way that the derivatives $\pa_z |\psi_1|$ and $\pa_z |\psi_2|$  vanish outside the region of interaction of the solitons as in Figs. 1 and 2. Notice that the analytic result computed in (\ref{ssf}) is time independent, whereas the numerical result (\ref{numdx}) is not. There are a number of sources of time dependence. First, the finite box approximation for the numerical calculation renders the orthogonality of the zero-mode $\pa_x \psi_1$ with the other vectors of the basis not to be  exact. Second, the factors $e^{iw t}$ which appear in the dynamics of the individual dark solitons and the time dependence of the numerically calculated solution $\widetilde{\Psi}(x,t)$. We extract the desired spatial shift by fitting the numerically calculated $\widetilde{\D x}(t)$ to  a straight line as 
\br
\label{fit}
\widetilde{\D x}(t)= \D x + (\D V) t, 
\er 
where $\D x$ will be the numerically estimated spatial shift of the black soliton, to be compared with the relevant analytical results in each case, i.e. expression (\ref{lim2}) for the NLS case and (\ref{shiftcqnls}) for the CQNLS case, respectively; $\D V$ represents the velocity change of the soliton. In general the last term in (\ref{fit}) accounts for the higher order [$(v\g)^{-n},\,n\ge 2$] time dependent terms which has not been considered in our analytical calculations. In the numerical simulations of the NLS and CQNLS models below we have verified that the velocity change behaves as $(\D V)  =0 + {\cal O} [(v\g)^{-2}]$. In general the higher order terms will arise when the contributions from $\D W_{1}h, \D W_{2}^1h$ and $\D W_{2}^2h$ in the r.h.s. of (\ref{eqh}) are properly taken into account, in addition to the source term $S$ which has already been considered in our calculations. However, we expect that in the integrable NLS model $\D V  =0$,  to all orders in the perturbation expansion. The relevant developments for the higher order terms will require the knowledge of a complete set of orthonormal basis $\{ u_{\l}(x)\}$ in (\ref{eigen}) which we postpone to a future publication.  

We performed extensive numerical simulations for the both the integrable NLS and the non-integrable cubic-quintic NLS models in order to check our results. The simulations are carried out by varying  the velocity of the incoming soliton while the other set of parameters of the model are maintained fixed, and afterwards by fixing the value ($v \g$) and varying the remaining constants of the model (e.g. $\{|\psi_0|, \a,\b\}$ in the CQNLS model). Next we present the numerical simulation results for the NLS model and the CQNLS models, respectively.

{\sl NLS model.} We have solved numerically the integrable NLS model defined in (\ref{nls2}). We develop this procedure for an initial condition ($t=0$)  suitable for the interaction of gray and  black solitons   
\br
\label{dark2}
\psi(x) &=& \frac{1}{\sqrt{\b}}\left\{ \begin{array}{ll}
 \{i v  + P_2  \tanh{[P_2  ( x + \g_2) ]} \}\, , &   -L < x < x_0\\
 P_1  \, e^{i \zeta} \,  \tanh{[P_1  (x + \g_1)]} \,  , &   x_0 < x < L
\end{array} \right. \\
\label{para11}
\zeta & \equiv & \arctan{\Big[ \frac{v}{P_2 \tanh{P_2 (\g_2+x_0)}}\Big]},\,\,\,\,\,P_1 \cosh{\Big[P_2 (\g_2+x_0) \Big]}= P_2 \cosh{\(P_1 (\g_1+ x_0)\)}\label{para22},
\er  
where the solution (\ref{dark111}) has been used, for the dark soliton one has $v \neq 0$ and for the black one  $v=0$. Despite the fact that one has an analytical solution for the gray-black system in the NLS model (\ref{2dark}), which describes their collision at any time, the initial condition (\ref{dark2}) considers the asymptotic form of each soliton when they are sufficiently far apart. The dark soliton is initially centered at $-\g_2$ and moves to the right with velocity $v $, the stationary black soliton is centered at $-\g_1$. Initially at the point $x_0$ one has $|\psi(x_0)|=|\psi_0|$ and the both constant phases of the solitons  match. The phase factor of the gray soliton $e^{i \zeta}$ in (\ref{para1}) has been introduced in order to match the constant phase in (\ref{dark2}). In the numerical simulation we will consider initially well-separated solitons, this amounts to considering the distance between them to be  several times the width of the solitons, i.e. $|\g_1-\g_2| > \frac{1}{P_2}> \frac{1}{P_1} $ . Notice that  the initial condition (\ref{dark2}) must satisfy the boundary condition (\ref{bc}) at $|x| \rightarrow L$ for $|\psi_0| = \frac{\sqrt{P_2^2 + v^2}}{\sqrt{\b }} = \frac{P_1}{\sqrt{ \b}}$. In the numerical simulation the domain considered is $D =[-L,L] $ with $L=20$, mesh size $h = 0.022$ and time step $\tau = 0.0003$. The domain $D $ is chosen such that the effect of the extreme regions near the points $x = \pm L$ do not interfere the dynamics of the solitons. In our numerical simulations we have used the so-called time-splitting cosine pseudo-spectral finite difference (TSCP)  method \cite{bao} suitable for the {\sl nvbc} and  the ${\sl cw}$ background at rest. The sound speed in the model (\ref{nlsd}) becomes $v_s  = \sqrt{\b} \, |\psi_0|$ and the gray solitons of the model (\ref{nlsd}) must propagate with velocity $|v| < v_s$.  While we do not show the numerical results as in Fig. 2, we have also checked the results for various values of the sound velocity $v_s$ reproducing accurately the relevant figures.  
 
\begin{figure}
\centering
\label{fig4}
\includegraphics[width=7cm,scale=3, angle=0, height=8cm]{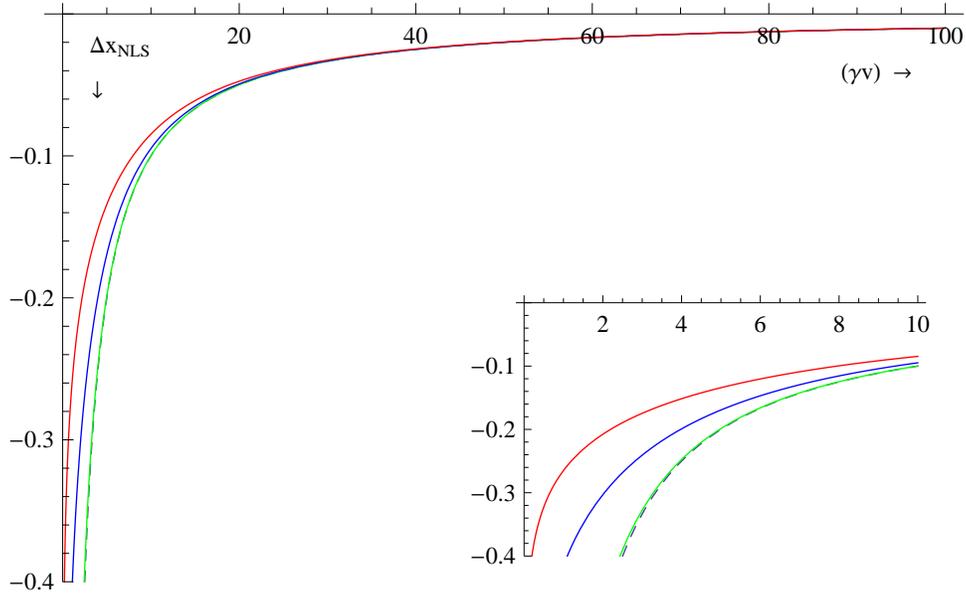}
\parbox{5in}{\caption{(color online) spatial shift $\D x$ as a function of $(\g v)$ in the integrable NLS model. The continuous lines correspond to the exact analytical result (\ref{analy}) for $v_s=1$(green),$v_s= 6$(blue) and $v_s=12$ (red), respectively. The theoretical prediction to first order of perturbation expansion (\ref{lim2}) is plotted as a dashed line. Notice that for velocities close to the speed of sound ($v\g >20$) it shows a good agreement with the analytical result for the values $v_s> 1$, whereas for low velocities it provides a good agreement  for $v_s \leq 1$ even for values $v \g \approx  3$. }}
\end{figure}
 
\newpage
 
{\sl Cubic-quintic NLS (CQNLS).} 

The numerical method TSCP is specialized in order to simulate the soliton collisions in the cubic-quintic NLS model (\ref{cqnls2}). As the 
 initial condition ($t=0$) suitable for the interaction of gray and  black solitons of the CQNLS model we consider the static solutions provided by (\ref{sl1})-(\ref{th1}) and (\ref{bl1})-(\ref{blp}), respectively, located far apart  
\br
\label{dark22}
\psi(x) &=& \left\{ \begin{array}{ll}
 \sqrt{\frac{\xi_{1} + r_2 \xi_{2} \tanh^2{[k_2 (x-x_2)]}}{ 1 + r_2 \tanh^2{[k_2 (x-x_2)]}}}\,e^{i \theta(x-x_2)}\,e^{i \zeta}, &   -L < x < x_0\\
 - \sqrt{ 
\frac{\xi \, r_1}{1+ r_1 \tanh^2{[k_1\, (x-x_1)]}}}\,\tanh{[k_1\, (x-x_1)]}\, e^{i \mu}\, e^{i \zeta}, &   x_0 < x < L
\end{array} \right. \\
\mu & \equiv & -\arctan{\sqrt{\frac{\xi_{2} r_ 2}{\xi_{1}}}};\,\,\,\,\, \theta(x-x_2) = -\arctan{\Big\{\sqrt{\frac{\xi_{2} r_ 2}{\xi_{1}}} \tanh{[k_2(x-x_2)]}\Big\}}\label{para22c}
\er  
The gray soliton is initially centered at $x_2$ and moves to the right with velocity $v = (\frac{\a \xi_{2} \xi_{1}}{3})^{1/2}$, the stationary black soliton is initially centered at $x_1$. Note that in the sound speed limit one has $r_2 \rightarrow 0$ (see eqs. (\ref{s22}) and  (\ref{limit22})), so the constant phase above vanishes $\mu \rightarrow 0$. Therefore it is necessary to introduce an overall factor $e^{i \zeta}$ in the initial condition (\ref{dark22}) in order to match the constant phase introduced for the black soliton (\ref{para1}). This parameter  also appears in the perturbative computation of the spatial shift in (\ref{phases1}). In the numerical simulation we will consider initially well-separated solitons, this amounts to considering the distance between them to be  several times the width of the solitons, i.e. $|x_1-x_2| > \frac{1}{k_2}> \frac{1}{k_1} $ . Initially one has  the conditions $|\psi(x_0)|=|\psi_0|, \pa_{x}|\psi(x)|_{x=x_0}=0\,$ and the both constant phases match at the point $x_0$. Moreover, the initial condition  satisfies the boundary condition (\ref{bc}) at $|x| \rightarrow L$, i.e.  $|\psi(|x|=L)|=|\psi_0|$. In the numerical simulation the domain considered is $D =[-L,L] $ with $L=30$ (sometimes $L=20$), mesh size $h = 0.022$ and time step $\tau = 0.0003$. The domain $D$ is chosen in order to give a well separated initial pulses and such that the effects of the extreme regions near the points $x = \pm L$ do not interfere the dynamics of the soliton interactions.  

\begin{figure}
\centering
\label{fig5}
\includegraphics[width=7cm,scale=3, angle=0, height=8cm]{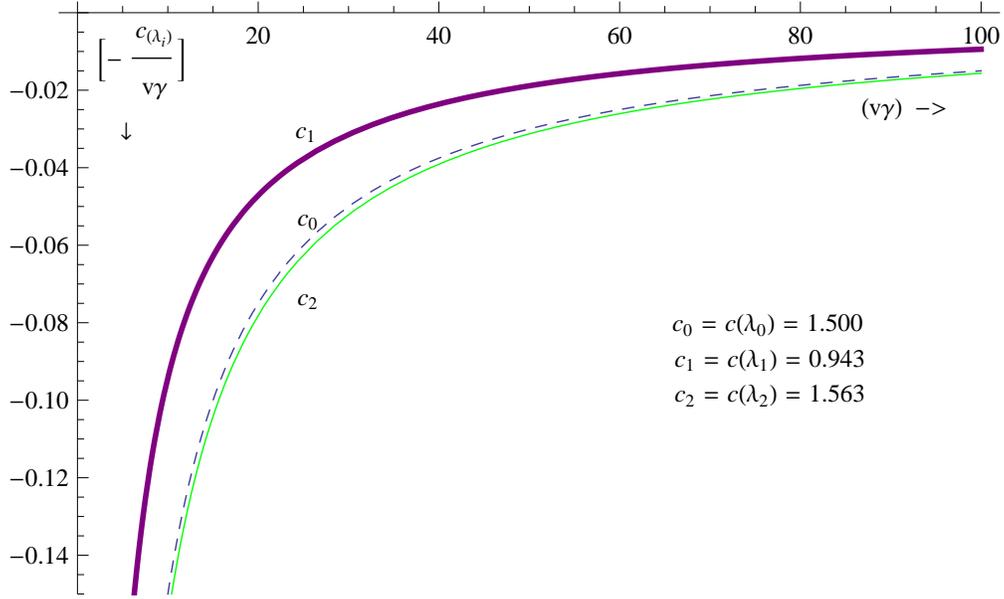}
\parbox{5in}{\caption{spatial shift $\D x$ as a function of $(\g v)$ in the CQNLS model. The theoretical prediction based on the perturbation theory(\ref{shiftcqnls}) is plotted for various values of the parameter $\l_0 = \{0, 0.526644\}$($c_0=1.5$), $\l_2 = 0.409785$($c_2=1.563$), $\l_1 = 0.62$($c_1=0.943$).}}
\end{figure}

\begin{figure}
\centering
\label{fig6}
\includegraphics[width=7cm,scale=3, angle=0, height=8cm]{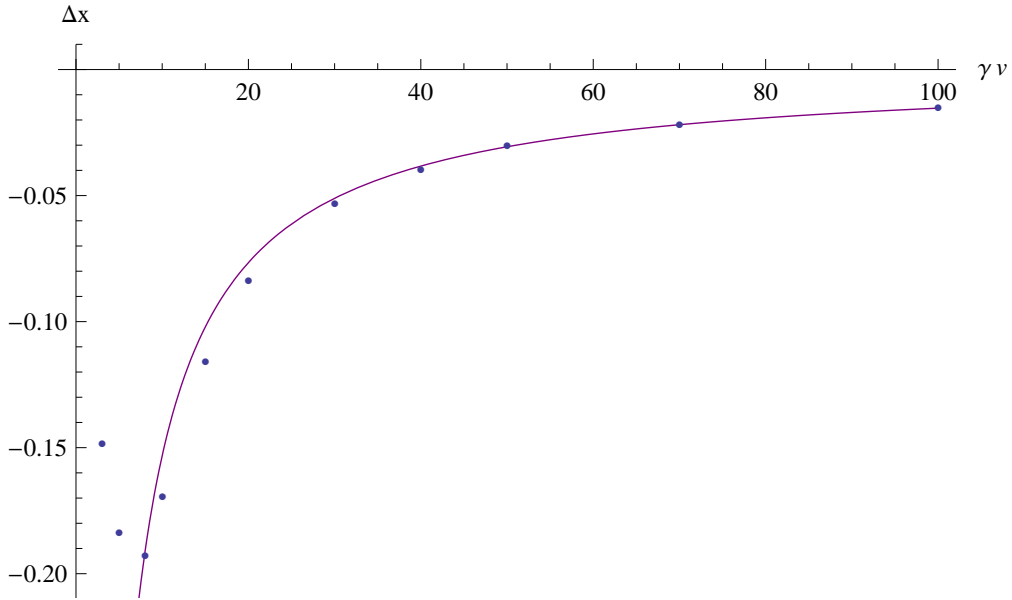}
\parbox{5in}{\caption{spatial shift $\D x$ as a function of $(\g v)$ in the CQNLS model. The theoretical prediction based on the perturbation theory result (\ref{shiftcqnls}) is plotted as continuous line  and the dots are the simulation results for the set of parameters $\a=0.2,\,|\psi_0|=2.75,\,\b=|\psi_0|^2=7.5625,\,v_s=6.76411,\,c(\l=0.2)=1.53164$.}}
\end{figure}
 
\begin{figure}
\centering
\label{fg6}
\includegraphics[width=7cm,scale=3, angle=0, height=8cm]{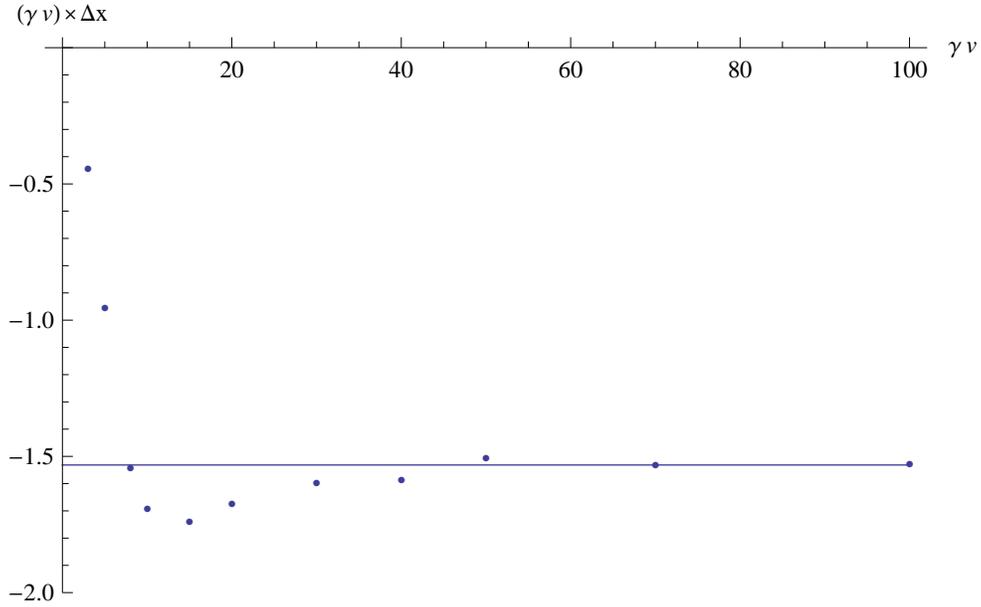}
\parbox{5in}{\caption{We plot the product $ (v \g) \D x$ to show how the analytic and numerical calculations behave as $(\g v)$ increases in the CQNLS model.}}
\end{figure}
 
\begin{figure}
\centering
\label{fg7}
\includegraphics[width=7cm,scale=3, height=8cm]{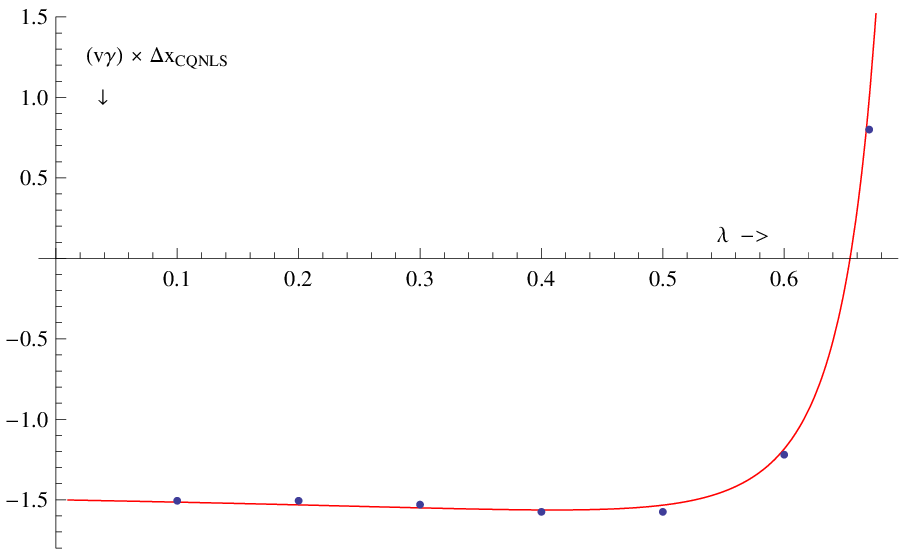}
\parbox{5in}{\caption{We plot the product $ (v \g) \times \D x_{CQNLS}$  as a function of $\l$ for  $(v\g)= 100$. The continuous line is the analytic result at leading order in $1/(v \g)$ and the dots correspond to numerical calculations. In the continuous line one has $(-1.5)$ at $\l =0$ and $\l=0.526644$, and the minimum value $(-1.56319)$ is located at $\l = 0.409785$. Notice that it changes sign  at $\l = 0.654221$.}}
\end{figure}
 
The Figs. 6, 7 and 8 show the comparison of the analytical result (continuous lines) for the spatial shift as presented in (\ref{shiftcqnls}) with the results obtained by numerical simulation (dots) of the gray-black soliton collision in the CQNLS model.   The Fig. 6 shows the behavior of the phase shift as the velocity of the incoming gray soliton is varied. In this figure we used $\l=0.2$ which corresponds to the set of values $|\psi_0|=2.75, \b=|\psi_0|^2= 7.5625$ and $v_s= 6.76411$. Notice that the leading order result (\ref{shiftcqnls}) reproduces the $v\g$ dependence of the spatial shift with a good approximation for high   velocities given by $(v\g ) > 20$. Moreover, the difference between the numerically calculated spatial shift and the one calculated using  (\ref{shiftcqnls}) diminishes as $(v\g) $ increases. The Fig. 7  shows how the analytic and numerical results approach each other as $(v\g)$ increases. In Fig. 8 we verify the $\l$ dependence of the spatial shift at a fixed value of $(v\g) =100$. This value for $(v\g)$ is fixed such that the analytical and numerical results have a good match for the particular set of values of the parameters chosen above. For the region $\l \in I_1=[0, 0.52]$ the numerical and analytic calculations are in good agreement. However, in order to achieve a good match in the region $\l \in I_2=<0.52, 0.75>$ we choose another set of parameters, some values  $|\psi_0| \le 2.5, \b=|\psi_0|^2$ have been used. Since the parameters we have chosen satisfy the relationship $\b=|\psi_0|^2$ one has $\l=\frac{\a |\psi_0|^2}{\b} \rightarrow \l = \a$. So by varying $\l$ one is effectively probing the spatial shift behavior as if the quintic coupling constant $\a$ was varying. Therefore, the value of $(v \g )$ and the set of values of $\l$ at which the analytic result (\ref{shiftcqnls}) provides a good agreement with the numerical calculations depend on the model parameters. Notice that the second interval $I_2$ contains the point $\l = 0.654221 \in I_2$ in which the analytic expression for the spatial shift changes sign (the interaction starts becoming attractive). Note also that at the limiting point  $\l=0.75$ the leading order result diverges. 

We checked through our numerical simulations that the velocity change behaves as $\D v = 0 + {\cal O} [(v\g)^{-2}]$. Finally, we conclude that the leading order calculation of the spatial shift is in good agreement with our numerical simulations for a wide range of parameter values.    
 
 \section{Discussions}
\label{sec6}

We developed a perturbative scheme to compute the spatial shift experienced by a black soliton after collision with an shallow dark soliton in deformed NLS models. The colliding shallow dark solitons pass through each other and for linearized perturbations the stationary soliton's perturbations can be expanded as a power series in the parameter $1/(v\g)$. This is possible due to the fact that  the overlapping region in space-time during collision possesses an area which is proportional to the factor $A_{int} \sim \frac{1}{v \g}$. This factor is extracted by simultaneous space-time and suitable field space transformations, which is provided for each deformed NLS model. We furnished a prescription to compute the leading order contribution to this free passage.   
 
We provided a closed expression for the leading order contribution to the spatial shift (\ref{ssf}). This expression depends on the the parameters of the theory, such as the coupling constants and the {\sl cwb} amplitude, and the function $f[I]$ which defines the equation of the deformed model (\ref{nlsd}).  Remarkably, this result does not depend on the knowledge of the explicit solutions of the dark solitons.  Two examples of applications are presented, the integrable NLS and the non-integrable CQNLS models. First, we have checked our analytical results with the integrable NLS model and found that they are in excellent agreement. Second, for the non-integrable CQNLS model we derived the leading order in ($1/(v\g)$) contribution (\ref{shiftcqnls}) which depends  on an effective parameter $\l$ and checked with simulations. We found a good match with the numerical simulation results for a wide range of the parameter values. 

It must be emphasized the role played by the constant parameters associated to the phase factors of the black soliton and the {\sl cwb} in the regularization procedure. This procedure has been performed in order to get finite values for the spatial shift in the both models presented above. 

Our analytical and numerical results show that qualitatively the shallow dark soliton collisions share some features with the kink-kink type collisions of the deformed SG model \cite{amin1}. In fact, the absolute value of the spatial shift $|\D x|$ decreases with increasing velocity in the both NLS and CQNLS models. In addition, in the CQNLS case and for fixed $(v\g)$ the spatial shift changes sign for certain  value  of the effective coupling constant $\l$. This resembles the behavior of the  spatial shift in terms of the deformation parameter $\a$ for the deformed SG with potential $V(\phi)=(1-\cos{\phi}) (1- \a \sin^2{\phi})$ \cite{amin1, amin2}.

It would be interesting to see the effects of higher order terms and the remaining  zero modes associated to phase, Galilean and scale invariance in future research. On the other hand, it is well known that the integrable models are characterized by the lack of velocity change after soliton collisions and that the quasi-integrable systems share some properties with their integrable counterparts. These properties of the quasi-integrable systems deserve future investigations in the framework of our formalism.

For a non-integrable theory with solitary wave solution there appear some  questions regarding the stability of the wave under small perturbations and  the properties of the solitary wave collisions. Most of the known results concern some calculations in the regimes which are close to either integrable or related to a particular relationship between the parameters of the colliding solitons, e.g.  high relative velocity, one of the solitons is significantly larger than the other one, fast thin solitons and slow broad solitons are among the cases considered in the literature. Related to these developments, when the wave configuration possesses a special space-time parity symmetry has been related to the concept of quasi-integrability \cite{jhep1}. In this context, the solitary wave interactions for the generalized KdV equation have been considered by Martel and Merle in a special regime \cite{martel}. We can also mention the work of G. Perelman on the bright solitary wave collisions of the deformed NLS  model (\ref{nlsd}) with focusing nonlinearity, where one soliton is small with respect to the other \cite{galina} and the asymptotic approaches to describe the evolution and collision of three waves of the generalized KdV model by Omel'yanov \cite{georgy}. 

\section{Acknowledgments}

The authors thank A.C.R. do Bomfim, H.F. Callisaya, L. da R. Mota, J. Parreira and A.M. Vilela for interesting discussions and Prof. L.A. Ferreira for discussions and hospitality at the IFSC-USP. The authors acknowledge FAPEMAT for financial support. The authors thank the anonymous referees for valuable comments and suggestions.

\appendix

\newpage

\section{Defocusing NLS: Exact analytic 2-dark soliton collision and spatial shift}
\label{dark}

Here we present the one and two-dark solitons of the NLS model. The spatial shift experienced by a black soliton after collision with a gray soliton is computed. Consider the defocusing NLS equation 
\br
\label{nlsdef}
i \pa_{t} \psi +  \pa_{xx} \psi - \beta |\psi|^{2} \psi=0,\,\,\,\, \beta >  0.
\er
The simplest solution of (\ref{nlsdef}) has the form of a continuous wave 
\br
\label{cw}
\psi = |\psi_0|  \exp{[i(k x + w t + x_0)]},
\er
provided that $w= -k^2/2 - \beta |\psi_0|^2$. In this work we will consider the continuous wave background at rest, i.e. $k=0$. The system (\ref{nlsdef}) is hyperbolic for small oscillations around $|\psi|= |\psi_0|$ and it has an associated sound speed given by $v_s = \sqrt{\beta} |\psi_0|$ \cite{chiron}. This  velocity plays an important role in the existence of solitary waves in the NLS model.  

The {\bf $1$-dark soliton} solution is given by \cite{jpa1} 
\br
\psi(x, t) &=&\frac{1}{\sqrt{\b}} \, e^{i w t}  \Big\{ \frac{1+ y\, \exp^{[2 P(x- v t - x_0)]}}{1+ \exp^{[2 P(x- v t - x_0)]}}\Big\},\,\,\,\, y = e^{2 i \eta},\,\,\,\,\eta=\arctan{(-P/v)} \\
&=&\frac{1}{\sqrt{\b}} \, e^{i w t} \Big\{ i v + P \tanh{[P (x - v t - x_0)]}\Big\}\label{dark111}
\er  
This solution possesses {\sl three} arbitrary real parameters, $v, P$ and $\b$. The intensity function $|\psi|$ moves at the velocity $v$, which is the velocity of the dark soliton. The dark soliton approaches constant amplitude $\sqrt{(P^2 + v^2)/\b} $ as $|x| \rightarrow \pm \infty$. As $x$ varies from $-\infty$ to $+\infty$ the soliton acquires a phase $2 \arctan{(v/P)}-\pi$. Moreover, at the center of the soliton, i.e. for $x-v t - x_0\equiv 0$, one has that the intensity becomes $|\psi|_{center} = v/\sqrt{\b}$. This center intensity is lower than the asymptotic amplitude $|\psi_0| $ and this property characterizes a dark soliton. Notice that this center intensity is controlled by the parameter $\eta$; i.e. this parameter defines the ``darkness'' of the soliton. The soliton with $v=0$ is the black soliton, whereas the moving soliton $v\neq 0$ is the so called gray soliton. The speed of the soliton must satisfy $|v|< v_s$.

The {\bf $2$-dark soliton} solution is given by \cite{jpa1}
\br
\label{2dark}
\psi(x,t)&=&|\psi_0| e^{i w t}\, \Big[\frac{1+ y_1 e^{ \G_1(x,t)}+ y_2 e^{ \G_2(x,t)} + r\,  y_1\, y_2\, e^{ \G_1(x,t)} e^{ \G_2(x,t)}}{1+ e^{ \G_1(x,t)}+ e^{ \G_2(x,t)} + r\, e^{ \G_1(x,t)} e^{ \G_2(x,t)}}\Big]
\\
\G_1(x,t) &\equiv&   2 P_1 (x-V_1 t - x_1);\,\,\G_2(x,t) \equiv  2 P_2 (x-V_2 t - x_2),\\
r &=& \frac{(P_1-P_2)^2 + (V_1-V_2)^2}{(P_1+P_2)^2 + (V_1-V_2)^2},\,\,\,y_j = e^{ i 2\d_j} ,\,\,\d_j= \arctan{(-P_j/V_j)},\\
&&-\pi/2 < \d_j < \pi/2,\,\,\,\,j=1,2. 
\er
The parameters satisfy
\br
P_1^2+ V_1^2 = P_2^2 + V_2^2 = \b |\psi_0|^2,\,\,\,\,\,\,w = -\b |\psi_0|^2  
\er

The $V_{j}$ are the soliton velocities and the $P_{j}$ are the parameters associated to the widths ( $\sim  \frac{1}{2 P_{j}} $) of each soliton. Next, let us compute the spatial shift experienced by a black soliton after collision with a gray soliton. Let $V_1=0$ be the black soliton velocity and $V_2=v>0$ the gray soliton velocity. Assume $P_1>0, P_2>0$ and that the gray soliton is initially located at $ x \rightarrow -\infty$ then one has $e^{\G_2} \rightarrow 0$, therefore from (\ref{2dark}) one can get $\psi(x,t_i) \sim   \Big\{ \frac{1+ y_1\, \exp{[2 P_1(x - x_1)]}}{1+ \exp{[2 P(x - x_1)]}}\Big\}$, which is the initial black soliton configuration at time $t_i$. Consider the gray soliton is  located at $x \rightarrow + \infty$ after it has already been collided with the black soliton, so $e^{\G_2} \rightarrow \infty $,  and from (\ref{2dark}) one has $\psi(x,t_f) \sim   \Big\{ \frac{1+ y_1\, \exp{[2 P_1(x - x_1+\frac{\log{r}}{2 P_1})]}}{1+ \exp{[2 P(x - x_1+\frac{\log{r}}{2 P_1})]}}\Big\}$. Comparing this with the initial configuration one has that the spatial shift experienced by the black soliton becomes \cite{zakharov}
\br
\D x &=&  \frac{1}{2 P_1} \log{r} \label{shiftnls1}\\
&=& \frac{1}{2 v_s} \log{\frac{\g -1}{\g+1}},\,\,\,\,\g \equiv \frac{1}{\sqrt{1-(v/v_s)^2}},\,\,\,v_s = \sqrt{\b} |\psi_0| \label{shiftnls2}
\er     

\section{Cubic-quintic NLS dark solitons}
\label{cq1}

In the following we will discuss some properties of the dark and black soliton type solutions of the cubic-quintic NLS model (CQNLS) defined in (\ref{cqnls2}). Substituting (\ref{para2}) into the eq. (\ref{cqnls2}) the corresponding expression in (\ref{ff2}) becomes
\br
\label{G2}
G[\phi_2] &=& \sqrt{\frac{\a}{3}} \frac{|\psi_0|^2-\phi_2^2 }{\phi_2} \sqrt{(\phi_2^2-\xi_{1})(\xi_{2}-\phi_2^2)}\\
\xi_{1} &=& \frac{(3 \b -2 \a |\psi_0|^2) - \sqrt{(3 \b -2 \a |\psi_0|^2)^2-12 v^2 \a}}{2 \a},\\
\xi_{2} &=& \frac{(3 \b -2 \a |\psi_0|^2) + \sqrt{(3 \b -2 \a |\psi_0|^2)^2-12 v^2 \a}}{2 \a},
\,\,\,\, \xi_{1} < |\psi_0|^2 < \xi_{2}.
\er
The set of parameters $\{ \xi_{1}, |\psi_0|, \xi_{2}\}$ are the roots of the functional $G[\phi_2]$ and plays an important role in characterizing some properties of the dark soliton. Integration of (\ref{ff2}) provides the dark soliton of the model \cite{crosta}
\br
\label{sl1}
\phi_2^2(z) &=& \frac{\xi_{1} + r_2 \xi_{2} \tanh^2{[k_2 (z-z_0)]}}{ 1 + r_2 \tanh^2{[k_2 (z-z_0)]}}\\
r_2 & \equiv & \frac{|\psi_0|^2-\xi_{1}}{\xi_{2}-|\psi_0|^2},\,\,\,k_2 \equiv \sqrt{\frac{\a}{3}} \sqrt{ (\xi_{2} - |\psi_0|^2) (|\psi_0|^2-\xi_{1})}\label{s22}.
\er
Notice that $k_2$ characterizes  the inverse soliton width and the root $\xi_{1}$ is the minimum intensity (dip) of the dark soliton. The maximum  intensity of the dark soliton approaches $\phi_2(\pm \infty)=|\psi_0|$. Moreover, for $v=0$ the small root $\xi_{1}$ vanishes and the dark soliton becomes a black soliton.  

Likewise using the solution (\ref{sl1}) in (\ref{ff3}) one has
\br
\label{th1}
\theta(z) = -\arctan{\Big[ \sqrt{\frac{\xi_{2}}{\xi_{1}}} \sqrt{r_2} \tanh{[k_2 (z-z_0)]}\Big]}+\theta_1,
\er
where $\theta_1$ is a constant of integration. For $\theta_1 = 0$ the phase of the dark soliton approaches the constant values $\theta^{\mp} = \pm  \arctan{\( \sqrt{\frac{\xi_{2} r_2}{\xi_{1}}} \)}$ as $z \rightarrow \mp \infty $, respectively,  and it is an odd function $\theta(-(z-z_0))=-\theta(z-z_0)$ in the moving frame of reference $(z, t)$. Therefore, the dark soliton becomes parametrized as in (\ref{para2}) with $\phi_2(z)$ given by (\ref{sl1}), $\theta(z)$ by (\ref{th1}) and  
\br
w = -\b |\psi_0|^2 + \frac{\a}{2} |\psi_0|^4
\er

In addition, the eq. (\ref{id1})   gives, upon integration and inversion the relationship 
\br
\label{relation1}
\phi^2_{2} = \xi_{2} \Big[ \frac{1+ \tan^2{\(\, \theta-\theta_1\)}}{ \frac{\xi_{2}}{\xi_{1}} + \tan^2{\(\, \theta -\theta_1 \) } } \Big]  
\er
   
The system (\ref{cqnls2}) is hyperbolic for small oscillations around $|\psi|= |\psi_0|$ and it has an associated sound speed given by $v_s = |\psi_0| \sqrt{f''[|\psi_0|^2]} = |\psi_0| \sqrt{\b - \a |\psi_0|^2}$ \cite{chiron}. Defining the parameter $
\g = 1/\sqrt{1-(\frac{v}{v_s})^2}$ one can write the sound speed $v \rightarrow v_s\, (\g >> 1)$ limit of the parameters as
\br
\label{limit22}
\xi_1 \rightarrow |\psi_0|^2,\,\,\,\,\xi_ 2\rightarrow \frac{3 v_s^2}{\a |\psi_0|^2}. 
\er

The stationary black soliton is obtained as the solution of the eq. (\ref{ff1}). So, one has  
\br
F[\phi_1] &=& \sqrt{\frac{\a}{3}} (|\psi_0|^2-\phi_1^2) \sqrt{\xi - \phi_1^{2}}\label{f111}\\
\phi^2_{1} &=& 
\label{bl1}
\frac{\xi \, r_1 \tanh^2{(k_1\, z)}}{1+ r_1 \tanh^2{(k_1\, z)}}\\
\label{blp}
\xi & \equiv &  \frac{1}{\a} (3 \b -2 \a |\psi_0|^2),\,\,\,r_1 \equiv \frac{|\psi_0|^2}{\xi -|\psi_0|^2},\,\,\,k_1=\frac{|\psi_0| \sqrt{\a} \sqrt{\xi -|\psi_0|^2}}{\sqrt{3}}.
\er
The minimum intensity of the static black soliton $\phi_1$ is zero and it approaches $|\psi_0|$ as $z \rightarrow \pm \infty$.  Moreover, the eqs. (\ref{d11})-(\ref{dx111}) supplied with  the eq. (\ref{f111}) give the energy of the black soliton of the CQNLS model
\br
\label{MCQ}
{\cal M}_{CQ} =\frac{1}{4}\sqrt{\frac{\a}{3}} \Big[|\psi_0| \sqrt{\xi -|\psi_0|^2}(\xi+2|\psi_0|^2)-\xi (\xi-4|\psi_0|^2) \arctan{(\frac{|\psi_0|}{\sqrt{\xi-|\psi_0|^2}})}   \Big]. 
\er
Notice that this soliton mass formula reproduces the NLS black soliton mass (${\cal M}_{NLS}=\frac{4}{3} \sqrt{\beta} |\psi_0|^{3}$) in the limit $\a\rightarrow 0$.

\end{document}